\documentclass[a4paper,11pt]{article}

\usepackage[a4paper,left=2.73cm,right=2.7cm,top=3cm,bottom=3.5cm]{geometry}

\usepackage{amsmath,amssymb,graphicx,caption,subcaption,enumerate,color,tikz}
\usetikzlibrary{arrows}
\usepackage{colortbl}
\usepackage{booktabs}

\definecolor{gray}{rgb}{.9,.9,.9}
\usepackage[colorlinks=true,linktocpage=true,linkcolor=blue,citecolor=blue]{hyperref}

\tikzstyle{block} = [rectangle, text width=5em, text  centered, rounded corners]
\tikzstyle{line} = [draw, very thick, -latex']
\tikzstyle{RGflow}= [->, shorten <=1pt, thick, dashed, color=black!70]

\addtocontents{toc}{\protect\setcounter{tocdepth}{2}}
\numberwithin{equation}{section}

\def\d{\mathrm{d}}
\newcommand{\dd}{\mathrm{d}}
\newcommand{\mt}[1]{\textrm{\tiny #1}}
\newcommand{\sac}{\, , \qquad}
\newcommand{\eqq}[1]{(\ref{#1})}
\def\cF{{\cal F}}

\newcommand{\ud}{U_\mt{dual}}
\newcommand{\uf}{U_\mt{flavor}}
\newcommand{\ust}{U_\mt{charge}}
\newcommand{\up}{U_\mt{pert}}
\newcommand{\uc}{U_\mt{cross}}

\newcommand{\Mq}{M_\mt{q}}
\newcommand{\um}{u_\mt{M}}

\newcommand{\ls}{\ell_s}
\newcommand{\gym}{g_\mt{YM}}

\newcommand{\gs}{g_{s}}

\newcommand{\Qc}{Q_\mt{c}}
\newcommand{\Qf}{Q_\mt{f}}
\newcommand{\Qst}{Q_\mt{st}}
\newcommand{\nq}{N_\mt{q}}
\newcommand{\nqp}{n_\mt{q}}
\newcommand{\nc}{N_\mt{c}}
\newcommand{\nf}{N_\mt{f}}
\newcommand{\nfp}{n_\mt{f}}

\newcommand{\ratio}{\rho}
\newcommand{\massratio}{m}

\newcommand{\bfunc}{{\cal B}}
\newcommand{\cfunc}{{\cal C}}
\newcommand{\At}{A_{t}}
\newcommand{\cB}{\mathcal B}

\newcommand{\be}{\begin{equation}}
\newcommand{\ee}{\end{equation}}
\newcommand{\bal}{\begin{align}}
\newcommand{\eal}{\end{align}}
\newcommand{\bse}{\begin{subequations}}
\newcommand{\ese}{\end{subequations}}

\newcommand{\td}{T_\mt{D6}}

\begin{document}

\begin{titlepage}

\thispagestyle{empty}

\begin{flushright}
\hfill{ICCUB-15-023}
\end{flushright}

\vspace{55pt}  
	 
\begin{center}

{\Huge \textbf{Three-dimensional super Yang-Mills \\ [4mm] with compressible quark 
matter}}
		
\vspace{30pt}
		
{\large \bf Ant\'on F. Faedo,$^{1}$ Arnab Kundu,$^{2}$ 
David Mateos,$^{1,\,3}$  \\ [2mm]
Christiana Pantelidou$^{1}$ and Javier Tarr\'\i o$^{1,\, 4}$}
		
\vspace{25pt}
		
{\normalsize  $^{1}$ Departament de F\'\i sica Fonamental and Institut de Ci\`encies del Cosmos,\\  Universitat de Barcelona, Mart\'\i\  i Franqu\`es 1, ES-08028, Barcelona, Spain.}\\
\vspace{15pt}
{ $^{2}$ Theory Division, Saha Institute of Nuclear Physics \\
1/AF Bidhannagar, Kolkata 700064, India.}\\
\vspace{15pt}
{ $^{3}$Instituci\'o Catalana de Recerca i Estudis Avan\c cats (ICREA), \\ Passeig Llu\'\i s Companys 23, ES-08010, Barcelona, Spain.}\\
\vspace{15pt}
{ $^{4}$
Universit\'e Libre de Bruxelles (ULB) and International Solvay Institutes,\\
Service  de Physique Th\'eorique et Math\'ematique, \\
Campus de la Plaine, CP 231, B-1050, Brussels, Belgium.}

\vspace{40pt}
		
\abstract{
We construct the gravity dual of three-dimensional, $SU(\nc)$ super Yang-Mills  theory with $\nf$ flavors of dynamical quarks in the presence of a non-zero quark density $\nq$. The supergravity solutions include the backreaction of $\nc$ color D2-branes and $\nf$ flavor D6-branes with $\nq$ units of electric flux on their worldvolume. For massless quarks, the solutions depend non-trivially only on the dimensionless combination $\rho=\nc^2 \nq / \lambda^2 \nf^4$, with $\lambda = \gym^2 \nc$ the 't Hooft coupling, and describe renormalization group flows between the super Yang-Mills theory in the ultraviolet and a non-relativistic theory in the infrared. The latter is dual to a hyperscaling-violating, Lifshitz-like geometry with dynamical and  hyperscaling-violating exponents $z=5$ and  $\theta=1$, respectively. If $\rho \ll 1$ then at intermediate energies there is also an approximate AdS$_4$ region, dual to a conformal Chern-Simons-Matter theory, in which the flow exhibits quasi-conformal dynamics.  At zero temperature we compute the chemical potential and the equation of state and extract the speed of sound. At low temperature we compute the entropy density and extract the number of low-energy degrees of freedom. For quarks of non-zero mass $\Mq$ the physics depends non-trivially on $\rho$ and $\Mq \nc/\lambda \nf$. 
}

\end{center}

\end{titlepage}

\tableofcontents

\hrulefill
\vspace{10pt}

\section{Introduction and discussion of results}
\label{intro}
Quantum Chromodynamics (QCD) at non-zero quark density is notoriously difficult to analyze. The only first-principle, non-perturbative tool, namely lattice QCD, is of very limited applicability due to the so-called sign problem \cite{deForcrand:2010ys}. It is therefore useful to construct toy models of QCD in which  interesting questions can be posed and answered. The gauge/string duality, or holography for short \cite{Maldacena:1997re,Gubser:1998bc,Witten:1998qj}, provides a framework in which the construction of some such models is possible. The goal is not to do precision physics but to be able to perform first-principle calculations that may lead to interesting insights applicable to QCD (see e.g.~\cite{Mateos:2011bs} for a discussion of the potential and the limitations of this approach). In the case of QCD at non-zero temperature, the insights obtained through this program  range from static properties to far-from-equilibrium dynamics of strongly coupled plasmas (see e.g.~\cite{CasalderreySolana:2011us} and references therein). 

A simple class of holographic models can be obtained from the supergravity solutions for a collection of $\nc$ D$p$-branes. In this case one finds an equivalence between the $d=p+1$ dimensional, $SU(\nc)$, supersymmetric gauge theory living on the stack of branes and string theory on the near-horizon limit of the geometry sourced by the D$p$-branes \cite{Itzhaki:1998dd}. Since the matter in these theories is in the adjoint representation of the gauge group, in order to consider a non-zero quark density new degrees of freedom in the fundamental representation must be included. On the gravity side this can be done by adding $\nf$ so-called flavor branes to the D$p$-brane geometry \cite{Karch:2002sh}. For conciseness  we will refer to these new degrees of freedom as `quarks' despite the fact that they include both bosons and fermions. Placing the theory at a finite quark density $\nq$ then corresponds to turning on $\nq$ units of electric flux on the flavor branes \cite{Kobayashi:2006sb}. This flux sources the same supergravity fields as a density $\nq$ of fundamental strings dissolved inside the flavor branes. We will thus refer to $\nq$ as the quark density, as the electric flux on the branes, or as the string density interchangeably.  Note that $\nc$ and $\nf$ are dimensionless integer numbers, whereas $\nq$ is a continuous variable with dimensions of (energy)$^{d-1}$.

If $\nf \ll \nc$ and $\nq$ is parametrically no larger than $\mathcal O(\nc^2)$ then there exists an energy range in which one can study this system in the so-called `probe approximation', meaning that the gravitational backreaction of the flavor branes and strings on the original D$p$-brane geometry can be neglected.\footnote{Since the tension of the strings is $\nc$-independent, their backreaction is of order $\nq/\nc^2$.}  However, for $d=3, 4$ the backreaction of the charge density always dominates the geometry sufficiently deep in the infrared (IR)  no matter how small $\nq$ is. In other words, the probe approximation always fails at sufficiently low energies \cite{Hartnoll:2009ns,Bigazzi:2013jqa}, and changing the value of $\nq$ simply shifts the energy scale at which this happens. Therefore, for $d=3, 4$ the inclusion of  backreaction  is not an option but a necessity in order to identify the correct ground state of the theory. 

In this paper we will find the fully backreacted supergravity solutions when $d=3$. As we will explain in Sec.~\ref{sec.setup}, we will distribute, or smear, the flavor branes and the strings  over the compact part of the geometry. We will focus on the case in which this geometry is an S$^6$, but  our results are also valid (with the sole modification of some numerical coefficients) if this is replaced by any other nearly-K\"ahler six-dimensional manifold, as in \cite{Faedo:2015ula}. On the gauge theory this  corresponds to replacing the maximally supersymmetric $SU(\nc)$ gauge  theory by an $\mathcal N =1$ quiver theory.

The $d=3$ case is simpler than the $d=4$ one because the gauge theory is asymptotically free even after the addition of flavor. In contrast, for $d=4$ the addition of flavor causes the gauge theory to develop a Landau pole in the UV. This does not preclude the study of the IR physics associated to the charge density, but the analysis is technically harder and hence it will be presented  elsewhere \cite{forth}.

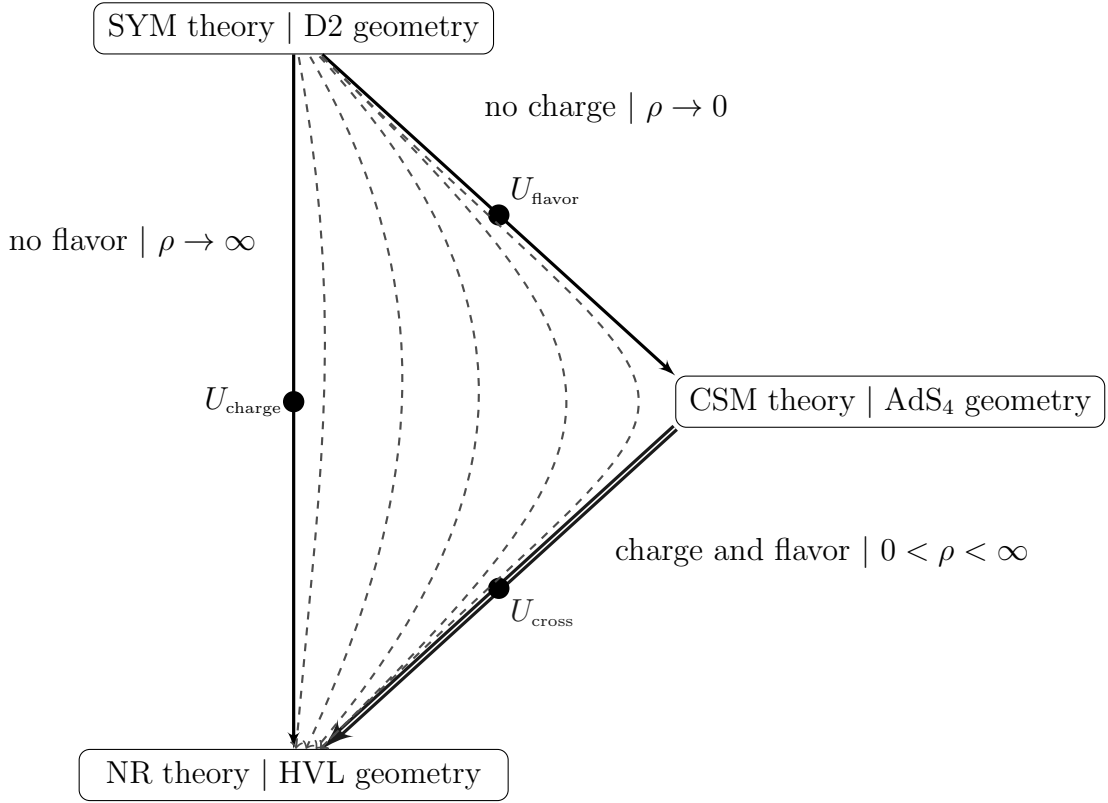
\begin{figure}[t!!!]
\begin{center}
\begin{tikzpicture}[auto]


\node [block, draw, text width=13em] (YM) {\large  SYM theory $|$ D2 geometry};
\node [below right of=YM, node distance=7cm] (dummy) {};
\node [block, draw,  right of=dummy, text width=14em, node distance=2.9cm] (CS) {\large CSM theory $|$ AdS$_4$ geometry};
\node [block, draw, below left of=dummy, node distance=7cm, text width=14em] (NR) {\large NR theory $|$ HVL geometry};


\path [line] (YM) -- node (nochargearrow) {\large $\uf$} (CS.north west);
\path [line] (YM) -- node (noflavorarrow)  [left]  {\large $\ust$} (NR);
\path [line, double, color=black!90] (CS.south west) -- node (conformalarrow) {\large   $\uc$} (NR);


\fill (noflavorarrow.east) circle [radius=4pt] node (noflavorcircle) { };
\fill (nochargearrow.south west) circle [radius=4pt] node (nochargecircle) { };
\fill (conformalarrow.north west) circle [radius=4pt] node (conformalcircle) { };


\node [above right of=nochargecircle, node distance=2cm] {\large no charge $|$ $\rho\to0$};
\node [above left of=noflavorcircle, node distance=3cm] {\large no flavor $|$ $\rho \to \infty$};
\node [below of=CS, node distance=2cm] { \hspace{-5em} \large charge and flavor $|$ $0< \rho < \infty$};


\draw [RGflow] (YM) to[in=85, out=280]  (NR);
\draw [RGflow] (YM) to[in=65, out=300]  (NR);
\draw [RGflow] (YM) to[in=50, out=312.5,looseness=1.2]  (NR);
\draw [RGflow] (YM) to[in=45, out=315,looseness=1.7]  (NR);
\draw [RGflow] (YM) to[in=45, out=315,looseness=2.2]  (NR);
\end{tikzpicture}
\caption{\small Pictorial representation of the family of supergravity solutions and corresponding RG flows  presented in this paper.}\label{fig.RGflows}
\end{center}\end{figure}

Many qualitative features of the solutions that we will construct can be understood by considering the limits in which the backreaction of the strings or the flavor branes dominate over one another. The supergravity solution in which the strings dominate was constructed in \cite{Faedo:2014ana} following the $d=4$ example of \cite{Kumar:2012ui}. It interpolates between the D2-brane geometry at large values of the holographic coordinate, and a hyperscaling-violating Lifshitz (HVL) solution at small values of the holographic coordinate. This interpolating solution is dual on the gauge theory to a Renormalization Group (RG) flow between the SYM theory in the ultraviolet (UV) and a non-relativistic (NR) theory in the IR, as one would expect from the presence of a charge density. This flow is represented by the vertical line on the left-hand side of Fig.~\ref{fig.RGflows}.

The HVL solution  has a four-dimensional metric of the form
\be\label{eq.HVLifmetric}
\d s^2 = \left( \frac{r}{L} \right)^{-\theta} \left[ - \left(\frac{r}{L} \right)^{2z} \d t_\mt{IR}^2 + \left( \frac{r}{L} \right)^2 \d x_2^2 + \beta^2 \left( \frac{L}{r} \right)^2 \d r^2 \right] \ ,
\ee
where $z$ is the dynamical exponent characterizing the different scalings of time and space, and $\theta$ is the HV exponent capturing the scaling properties of the line element. 
 
The normalization constant $\beta$ is arbitrary from the viewpoint of the IR geometry, but it takes a specific value when this geometry is connected to the UV geometry  \eqref{eqn.D2branes} along an RG flow. In our solutions the dynamical and HV exponents take the values
\be
\label{exp}
z=5 \,, \qquad \theta =1 \,,
\ee
respectively. The emergence of a Lifshitz geometry \cite{Kachru:2008yh} with HV in the IR could have been expected based on the analysis of \cite{Charmousis:2010zz,Gouteraux:2011ce}. 

The existence of this flow shows that the far UV is unmodified by the charge density. Indeed, in this limit the backreaction of the charge falls off as 
\be
\label{chargefalloff}
\ust^4/U^4 \,,
\ee
with \be\label{ucharge}
\ust \sim \lambda^{1/2} \, \left( \frac{\nq}{\nc^2} \right)^{1/4} 
\ee
and $\lambda = \gym^2 \nc$  the 't Hooft coupling constant of the gauge theory, which has dimensions of energy. The holographic coordinate $U$, which we will often use throughout this paper, also  has dimensions of energy and therefore it can be directly identified with an energy scale in the gauge theory. We thus see that $\ust$ is the energy scale at which the presence of the charge produces an effect of order one on the gauge theory state, or equivalently the scale at which  the backreaction of the strings  becomes of order unity on the gravity side. 

The supergravity solution with flavor branes and no strings was constructed in \cite{Faedo:2015ula}. In this case the gauge theory is $d=3$ SYM coupled to $\nf$ flavors of dynamical quarks but no charge density is turned on. As depicted by the top diagonal line in Fig.~\ref{fig.RGflows}, the supergravity solution interpolates between the D2-brane geometry in the UV and an $\text{AdS}_4\times \text{S}^6$ geometry in the IR. In the gauge theory this corresponds to an RG flow between SYM in the UV and an IR fixed point described by a conformal Chern-Simons-Matter (CSM) theory. As with the charge density, the presence of flavor leaves the UV unmodified, 
since its effect  falls off as 
\be
\label{flavorfalloff}
\uf/U 
\ee
with 
\be
\label{uflavor}
\uf\sim \lambda \frac{\nf}{\nc} \,,
\ee
but  it produces an effect of order unity at the energy scale $\uf$. Below this scale the geometry is approximately  $\text{AdS}_4\times \text{S}^6$ and the flow approaches the IR fixed point. 

The natural question that arises is what happens if a non-zero charge density is added to the CSM theory. The answer can be guessed by noting, as we will show, that the addition of strings to the AdS$_4$ geometry is a relevant deformation from the viewpoint of this fixed-point geometry. In other words, the strings leave the UV of the AdS$_4$ geometry unmodified but they induce a flow to a different geometry in the IR. Complementarily, we will show that the addition of flavor branes to the HVL geometry corresponds to adding an irrelevant deformation to the dual NR theory. This means that the flavor leaves the IR of the HVL geometry intact but modifies the UV of this geometry. Therefore, it is natural to expect  that the addition of strings to the AdS$_4$ geometry triggers a flow that drives the solution to an HVL solution in the IR, thus closing the triangle in Fig.~\ref{fig.RGflows} with the bottom diagonal line. Note that, because the flow starts at a fixed point with no dimensionful scales, all values of $\nq$ are physically equivalent since $\nq$ simply sets the units along the flow.

As we will see, if both flavor branes and strings are present from the beginning, the full set of supergravity solutions depending on $\nc, \nf, \nq$ and  $\lambda$ can be reduced through a simple rescaling to a family parametrized only by the dimensionless ratio of the two scales introduced above:
\be
\ratio \sim 
\left( \frac{\ust}{\uf} \right)^4 \sim 
\frac{1}{\lambda^2}  \left( \frac{\nc}{\nf} \right)^4 \frac{\nq}{\nc^2} \ , 
\label{rhoratio}
\ee
where the fourth power is chosen for  convenience. In other words, physical observables in any two solutions with arbitrary values of $\nc, \nf, \nq$ and $\lambda$ but with the same value of $\rho$ are related to one another by a rescaling. In this sense, we may say that the physics depends on $\nc, \nf, \nq$ and $\lambda$ nontrivially only through $\rho$.

Several flows with different values of $\rho$ are depicted in Fig.~\ref{fig.RGflows} by the dashed curves. For any value  $\rho > 0$ the flow begins in the UV at the SYM theory and ends in the IR at a NR theory. On the gravity side the solution starts with the D2-brane geometry at large radius and ends with an HVL geometry in the IR. We emphasize that the precise NR theory to which the theory flows depends on the value of $\rho$. Although in all cases the dynamical and the HV violating exponents are given by \eqq{exp}, other features are different. A simple example is the number of active degrees of freedom at low temperature, as measured by the entropy density. The exponents \eqq{exp} imply that this must  scale as 
\be
\label{prop}
s \sim c \, T^{(2-\theta)/z} \sim c \, T^{1/5} \,,
\ee
but the $T$-independent constant $c$ is not fixed by the scaling properties of the solution. We will see in Sec.~\ref{entropysection} that this constant, which  measures the number of low-energy degrees of freedom, depends on the values of $\nc, \nf, \nq$ and $\lambda$. 

In the limit $\rho\to \infty$ the flow is dominated by the backreaction of the strings: the scale $\ust$ becomes much larger than $\uf$, meaning that the flow is driven by the charge to the HVL fixed point well before the flavor can have a significant effect. Note from \eqq{chargefalloff} and \eqq{flavorfalloff} that the backreaction of the charge falls off with $U$ faster than that of the flavor. This means that, for any large but finite $\rho$, there is always a UV scale above which the flavor backreaction is larger than the charge backreaction. However, this is irrelevant because this happens in a region in which they are both a small effect with respect to the D2-brane geometry, and moreover this UV scale is pushed to infinity as $\rho \to \infty$. 

In the opposite limit, $\rho\to 0$, the hierarchy of scales is inverted and the flavor drives the flow to the AdS$_4$ fixed point before the charge can have a significant effect. If $\rho$ is very small but non-zero, then the flow is first driven very close to the AdS$_4$ fixed point but it eventually `realizes' that the charge density is non-zero and is then  driven to the HVL geometry in the deep IR. The scale $\uc$ at which the transition between AdS$_4$ and HVL takes place can be determined from the position in the holographic direction where the backreaction of the strings on the AdS$_4$ geometry becomes of order one. The result is
\be
\label{ucross}
\uc \sim \lambda^{3/7}\, \left( \frac{\nc}{\nf}  \right)^{1/7} \, \left( \frac{\nq}{\nc^2} \right)^{2/7} \,. 
\ee
Note that this scale is parametrically smaller than $\uf$ if $\rho \ll 1$, since 
\be
\uf^4 \sim \rho^{-1} \, \ust^4 \sim \rho^{-8/7}\, \uc^4 \,. 
\label{hi}
\ee
Thus in the small-$\rho$ limit the theory exhibits quasi-conformal or `walking' dynamics in the energy range $\uc \ll U \ll \uf$. In Sec.~\ref{quasi} we will verify this  explicitly by showing that the energy between a pair of external quarks separated a distance $L$ from one another exhibits the conformal scaling $E_{q\bar q}\sim 1/L$ in the energy range above.

\begin{figure}[t!!!]
\begin{center}
\begin{tikzpicture}[auto]

 \fill[red!25] (0,2.5-16/7)  -- (4,2.5) -- (8.5,4.75) -- (8.5,0) -- (0,0);
 \fill[cyan!20] (0,5) -- (0,2.5)  -- (4,2.5) -- (8.5,4.75) -- (8.5,5) ;
 \fill[orange!30] (0,2.5)  -- (0,2.5-16/7) -- (4,2.5);

 \draw [->] (-0.25,0) -- (8.75,0) node [below left]  {\large $\log \ratio$};
 \draw [->] (4,0) -- (4,5.5) node [above] {\large $\log U$};
 \draw  (4,-3pt) node [below] {\large $0$} -- (4,3pt);
 
 \draw (1.1,1.66) node {\large CSM theory};
 \draw (4.1,3.7) node {\large SYM theory};
 \draw (5,1) node {\large non-relativistic theory};

 \draw [very thick] (0,2.5-16/7) -- (4,2.5);
 \draw [very thick] (4,2.5) -- (8.5,4.75);
 \draw [very thick] (0,2.5) -- (4,2.5);

\end{tikzpicture}
\caption{\small Dynamics of the system at different energy scales $U$ as a function of the dimensionless ratio \eqq{rhoratio}. We choose a normalization such that the scale governing the SYM-CSM crossover is $\uf\sim \ratio^0$. The scale governing the YM-non relativistic crossover is then $\ust\sim\ratio^{1/4}$, and the scale governing the CSM-non relativistic crossover is $\uc\sim\ratio^{2/7}$. We emphasize that the lines separating the different regions simply indicate smooth crossovers between different asymptotic behaviors.}\label{fig.dynamics}

\end{center}\end{figure}
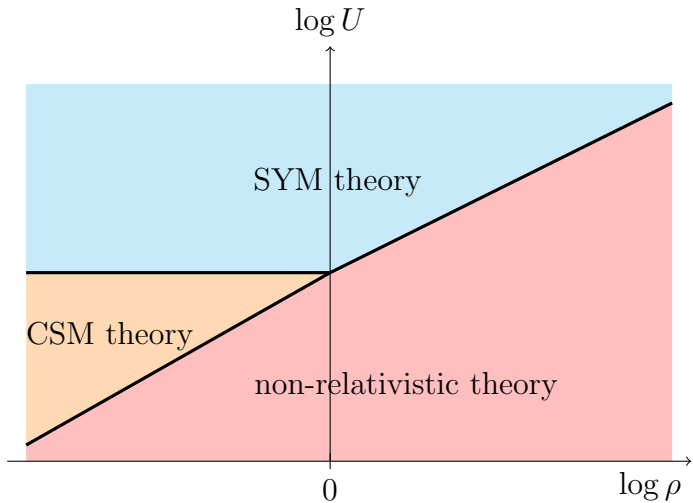

The dynamics of the theory at different energy scales for different values of $\rho$ is summarized in Fig.~\ref{fig.dynamics}.

This paper is organized as follows. In Sec.~\ref{fixedpoints} we describe the three geometries corresponding to the vertices of the triangle in Fig.~\ref{fig.RGflows}. In Sec.~\ref{crossovers} we determine the crossover scales between these geometries. In Sec.~\ref{full} we provide the full numerical solutions describing the RG flows depicted in Fig.~\ref{fig.RGflows}. In Sec.~\ref{quasi} we compute the energy of a quark-antiquark pair as a function of their separation and prove that, for RG flows with small $\rho$,  it exhibits the expected quasi-conformal behavior at intermediate separations. In Sec.~\ref{thermo} we compute the chemical potential, the equation of state and  the speed of sound at zero temperature, as well as the entropy density at low temperature.  Finally, in  Sec.~\ref{massive} we explain how the set of RG flows summarized in Fig.~\ref{fig.RGflows} is modified upon the introduction of a non-zero quark mass. 

\section{Fixed points}
\label{fixedpoints}
We will now study the solutions corresponding to each of the three vertices in the triangle of Fig.~\ref{fig.RGflows} and derive the crossover scales between them.  This analysis does not require the knowledge of the entire RG flows between the vertices. 
We refer to these solutions as `fixed points' because the D2 and the HVL solutions are dual to the UV and the IR fixed points of any flow, and the  AdS$_4$ solution can be approached arbitrarily closely along  RG flows with sufficiently 
small values of $\rho$. 

\subsection{Asymptotic D2-brane solution}
Maximally supersymmetric, three dimensional, $SU(\nc)$ SYM theory arises in type IIA string theory as the field theory on the worldvolume of a stack of $\nc$ D2-branes in the decoupling limit, $\ls\to0$, with fixed Yang-Mills coupling
\be
\gym^2=\gs/\ls \ .
\ee
In string frame, the  dual supergravity solution takes the form \cite{Itzhaki:1998dd} 
\bse\label{eqn.D2branes}\bal
\d s^2 & = \left( \frac{u}{L} \right)^{5/2} \d x_{1,2}^2 + \left( \frac{L}{u} \right)^{5/2} \left( \d u^2 + u^2\, \d \Omega_6^2 \right) \ , \\
e^\phi & =  \left( \frac{L}{u} \right)^{5/4} \ , \\[2mm]
F_6 & = \Qc \,\omega_6 \ ,\label{eqn.D2sRR}
\end{align}\ese
where $\d \Omega_6^2$ is the metric of a unit-radius six-sphere with volume form $\omega_6$ and volume \mbox{$V_6=\int \omega_6$}. The equations of motion relate the D2-brane charge $\Qc$ to the length scale $L$ as $\Qc = 5L^5$. Moreover, in the quantum theory the Ramond-Ramond (RR) flux must  satisfy the quantization condition
\be
\frac{1}{(2\pi\ls)^5\,\gs} \int F_6 = \nc \ ,
\label{quant}
\ee
which relates the dimensionful constants $\Qc$ and $L$ to the number of colors of the gauge theory
\be\label{eqn.Ncquantization}
\Qc = 5 L^5 = \frac{(2\pi \ls)^5}{V_6} \, \gs\,\nc \ .
\ee
Note that, without loss of generality, the dilaton in \eqq{eqn.D2branes} is normalized in a $g_s$-independent way, so that the local string coupling is actually $g_s e^\phi$. Consistently, we have included an explicit factor of $g_s$ in \eqq{quant}.

The holographic coordinate $u$ in \eqq{eqn.D2branes} has dimensions of length.  As explained in \cite{Itzhaki:1998dd}, it is useful to define a new radial coordinate 
\be
\label{bigu}
U = u/2 \pi \ell_s^2
\ee
with dimensions of energy which can be directly identified with an energy scale in the gauge theory. The solution \eqq{eqn.D2branes} provides a valid description in the range in which both the curvature in string units and the dilaton are small. This region is given by 
\be
\label{2}
\ud \ll U \ll \up \ ,
\ee
with 
\be
\label{pertdual}
\ud \equiv \lambda\, \nc^{-4/5} \sac
\up \equiv \lambda 
 \,,
\ee
and is parametrically large in the large-$\nc$ limit.

\subsection{Hyperscaling-Violating Lifshitz solution}
Adding a quark density to the SYM theory above is dual on the gravity side to adding a density of strings appropriately distributed. In the UV this induces a subleading correction to the solution \eqq{eqn.D2branes} that falls off as in \eqq{chargefalloff}. In the IR, this drives the theory to an HVL geometry \cite{Faedo:2014ana}. Here we summarize the necessary results for the discussion of the crossover scales; additional details will be presented below. 

The metric and dilaton are given by 
\bse\label{eqn.HVLifshitz}\bal
\d s^2 & = \frac{38416}{6655\, \left(\Qst\,L\right)^2} \left[ - C_\mt{t}   \, \frac{r^{11}}{L^{11}}  \d t^2 +  \frac{r^3}{L^3}  \d x_2^2 + \frac{363}{343} \, \frac{L}{r} \,\Qst\,L  \left(  \d r^2 + \frac{5}{12} \, r^2 \d \Omega_6^2 \right) \right] \ ,  \\[2mm]
e^\phi & =\frac{2^5\, 7^3}{11^3} \, \frac{\sqrt{7/5}}{\left( \Qst \, L \right)^{5/2}}\,  \frac{r^{5/2}}{L^{5/2}}  \ ,
\end{align}\ese
where $L$ is the same as in \eqref{eqn.D2branes}, $C_\mt{t}$ is a positive number (determined dynamically by the RG flow from the UV) that we will discuss further in Sec.~\ref{entropysection}, and the constant 
\be
\Qst  = 2\pi \ls^2\, \lambda\, \frac{\nq}{  \nc^2} 
\ee
measures the string charge density. The precise numerical factors and the powers of $\Qst$  in \eqref{eqn.HVLifshitz} are such that the  reduced four-dimensional metric is of the form given in Eq.~\eqref{eq.HVLifmetric} with IR time $t_\mt{IR}=\sqrt{C_\mt{t}}\, t$, and coincides exactly with the solution given in  Ref.~\cite{Faedo:2014ana}. In order to compare with the notation in that paper,  the reader should note that
\be
\Qst\big|_\text{here} =\frac{Q}{L}\Big|_\text{there}  
\,.
\ee

The dynamical and HV exponents of the solution are given by \eqq{exp} and 
can be read off from the fact that, in Einstein frame, the metric behaves as
\be
\d s^2 \sim r^{-2\theta/D} \Big[ - r^{2z} \d t^2 + r^{2} \d x_2^2 + r^{-2} \left( \d r^2 + r^2 \d \Omega_6^2 \right) \Big] \,,
\ee
with $D=8$ the number of spatial directions excluding the holographic direction.\footnote{Reducing upon the S$^6$ and using $D=2$ yields a metric like \eqref{eq.HVLifmetric} with the same exponents \cite{Perlmutter:2012he}.}

The metric and the dilaton \eqq{eqn.HVLifshitz} are supported by the RR fluxes
\be
F_6 = \Qc \, \omega_6 = 5\,L^5\, \omega_6 \ , \qquad F_2 = \Qst \, \d x^1 \wedge \d x^2 \ ,
\label{ff}
\ee
associated to the presence of a stack of D2-branes and a density of strings, respectively. In order to ensure that the string density of the perturbed D2-brane solution in the UV and the HVL solution in the IR are the same, as it should be for them to be part of the same RG flow, we must require that the $x$-directions are normalized in the same way. This can be accomplished by matching the IR and the UV forms of the $g_{xx}$ components of the effective, S$^6$-reduced metric. This leads to the following identification between the IR coordinate used in \eqq{eqn.HVLifshitz} and the UV coordinate used in \eqq{eqn.D2branes} \cite{Faedo:2014ana}: 
\be\label{eqn.HVLifD2radialchange}
\frac{r}{L} = \left( \frac{u}{L} \right)^{7/2} \,.
\ee

The range of validity of the solution \eqq{eqn.HVLifshitz}, namely the region where 
both the dilaton and the curvature in string units are small, is given by 
\be
\label{1}
U_\mt{lower} \ll U \ll U_\mt{upper}
\ee
with 
\be
\label{uplow}
U_\mt{lower} = \lambda^{3/7} \left( \frac{\nq}{\nc^2} \right)^{2/7} \sac 
U_\mt{upper} = \nc^{4/35} \, U_\mt{lower} \,,
\ee
which is  parametrically large in the large-$\nc$ limit. 

In order for the region where the geometry transitions between the D2-brane solution and the HVL solution to be described by type IIA supergravity, we must require that both \eqq{2} and \eqq{1} be satisfied at $U=\ust$, which translates into the following bounds on the charge density:
\be\label{eqn.chargeconstraints}
\lambda^2\, \nc^{-16/5}  \ll \frac{\nq}{\nc^2} \ll \lambda^2 \ .
\ee
Note that both bounds are compatible with $\nq$ being of order $\nc^2$. If the lower bound is relaxed, then there is a region along the flow where the dilaton is large and the correct description of that region is provided by eleven-dimensional supergravity.

\subsection{AdS$_4$ solution}
Adding $\nf$ flavors of dynamical quarks to the three-dimensional SYM theory  is dual on the gravity side to adding $\nf$ D6-branes. Ref.~\cite{Faedo:2015ula} constructed the corresponding solution when the D6-branes are smeared over the S$^6$ directions transverse to the D2-branes in a way that restores the maximum amount of symmetry compatible with $\mathcal N=1$ supersymmetry. In the UV this induces a subleading correction to the solution \eqq{eqn.D2branes} that falls off as in \eqq{flavorfalloff}. In the IR, this drives the theory to a fixed point described by an $\text{AdS}_4\times \text{S}^6$ geometry. Here we summarize the necessary results for the discussion of the crossover scales; further details will be given below. 

The construction of \cite{Faedo:2015ula} makes use of the fact that the S$^6$ admits a nearly K\"ahler (NK) structure, which implies that it is equipped with 
a real two-form, $J$, and a complex three-form, $\Omega$, satisfying:
\bse\label{eqn.SU3structure}\bal
\d J = 3 \,\text{Im}\,\Omega \ , &\qquad \d \,\text{Re}\,\Omega = 2\,J\wedge J \ , \\
J \wedge \Omega=0 \ , &\qquad \frac{1}{6} \, J \wedge J \wedge J = \frac{i}{8} \Omega \wedge \overline{\Omega} = \omega_6 \ , \\
*_6 J = \frac{1}{2} J\wedge J \ , &\qquad *_6 \Omega = -i \Omega \ .
\end{align}\ese
A specific construction of $J, \Omega$  in a particular coordinate system was given in \cite{Faedo:2015ula}.  These forms provide a natural basis for writing  the RR fluxes of the solution,
\be
F_6 = \Qc \, \omega_6 = 5 \, L^5 \, \omega_6 \ , \qquad F_2 = \Qf \, J \,,
\ee
where for massless quarks one has a constant
\be\label{eqn.Qfdef}
\Qf = \frac{2\pi\ls^2}{V_2} \,\, \lambda\, \frac{\nf}{\nc} \,,
\ee
with $V_2=\int J$, measuring the number of D6-branes. Alternatively we can view $\Qf$ as characterizing the strength of the backreaction of the flavor branes on the color geometry. The RR six-form is the same one as in Eqn.~\eqref{eqn.D2sRR}, originating from the stack of D2-branes, whereas the RR two-form violates the Bianchi identity, $\d F_2\neq 0$, signalling the presence of D6-brane sources.  The fluxes above support the following metric and dilaton  
\bse\label{eqn.AdS}\bal
\d s^2 & = \frac{R^2}{L_\mt{IR}^2} \d x_{1,2}^2 + \frac{L_\mt{IR}^2}{R^2} \left( \d R^2 + \frac{9}{4} R^2 \d \Omega_6^2 \right) \ , \\
\gs\, e^\phi & =  \left( \frac{2^{17}}{5^4}\,\frac{\pi^2}{3} \, \frac{\nc}{\nf^5} \right)^{1/4} \ ,
\end{align}\ese
where  
\be
L_\mt{IR}  = \ls \left( \frac{2}{3^5\,\pi^2} \, \frac{\nc}{\nf} \right)^{1/4} 
\ee
is the AdS$_4$ radius. The requirement that type IIA supergravity provides a reliable description of this geometry implies that 
\be
\nc^{1/5} \ll \nf \ll \nc \ ,
\ee
which in turn guarantees that 
\be
\ud \ll \uf \ll \up \,, 
\ee
with $\ud, \up$ defined as in \eqq{pertdual}. This means that, in fact, the entire flow from the D2-geometry to the AdS$_4$ geometry is well described by type IIA supergravity.

\section{Crossover scales}
\label{crossovers}
With the three solutions \eqq{eqn.D2branes}, \eqq{eqn.HVLifshitz} and \eqq{eqn.AdS} in hand, we are now ready to show how the crossover scales \eqq{ucharge}, \eqq{uflavor} and \eqq{ucross} are determined. As mentioned in Sec.~\ref{intro}, $\ust$ and $\uf$ can be determined from the full solutions interpolating between the D2-geometry and the HVL and AdS geometries, respectively. These solutions are known explicitly enough to determine the radial positions at which the backreaction of the strings \cite{Faedo:2014ana} or of the flavor branes \cite{Faedo:2015ula} becomes of order unity. Translating between these positions and gauge theory energy scales leads to \eqq{ucharge} and \eqq{uflavor}.

Alternatively, one may determine the three crossover scales by equating the dilatons in the three solutions \eqq{eqn.D2branes}, \eqq{eqn.HVLifshitz} and \eqq{eqn.AdS}. This amounts to identifying the crossover point with the point at which the dilaton changes behavior from one solution to another. Equating the dilaton in \eqq{eqn.D2branes} and \eqq{eqn.HVLifshitz} and using \eqq{eqn.HVLifD2radialchange} to express both dilatons in terms of the same radial coordinate $u$, and then translating this into an energy scale via \eqq{bigu}, leads to $\ust$. Equating the dilatons in \eqq{eqn.D2branes} and \eqq{eqn.AdS} and using again \eqq{bigu} leads to $\uf$. Finally, equating the dilatons in \eqq{eqn.HVLifshitz} and \eqq{eqn.AdS}, using \eqq{eqn.HVLifD2radialchange} to translate from $r$ to $u$, and then using \eqq{bigu} to translate from $u$ to $U$, leads to $\uc$.

We will now show that the three scales above can also be determined in a third way, namely by comparing the stress tensor of a D6-brane probe with strings dissolved in it with the stress tensor supporting the geometry in which the D6-brane is placed. The strategy is to determine the radial position at which the stress tensor of the brane-plus-strings system  would compete with that of the background geometry. At that point the backreaction would become of order unity. This will also allow us to show in which geometries the addition of flavor branes or strings corresponds  to a relevant or irrelevant deformation on the field theory side. 

For the purposes of this discussion it suffices to consider the Dirac-Born-Infeld (DBI) part of the D6-brane action (see the paragraph below Eqn.~\eqq{effect} for the justification). This takes the form
\be
\label{single}
S_\mt{DBI} = - \td \int \d^7\xi \,e^{-\phi} \sqrt{-{\cal P}[G]+ \mathcal F }
\ee
with 
\be
\td= \frac{1}{(2\pi \ell_s)^6 \,g_s \ell_s}  
\ee
the tension of the brane,  ${\cal P}[\cdots]$ the pullback of a spacetime object in string frame to the worldvolume of the brane, 
\be
\mathcal F = {\cal P}[B] + 2\pi\ls^2\, \d A \,,
\ee
$B$ the Neveu-Schwarz (NS) two-form and $A$ the BI gauge potential. We assume that the probe brane wraps an equatorial $\text{S}^3\subset\text{S}^6$ and is extended along the Minkowski and radial directions. The presence of strings dissolved inside the brane is encoded in an electric BI potential of the form
\be
\label{at}
A=\At(y) \d t \ ,
\ee
with $y$ the holographic radial coordinate.
For backgrounds of the form 
\be\label{eqn.metricansatz2}
\d s^2 = G_{tt}(y)\, \d t^2 + G_{xx}(y)\, \d x_2^2 + G_{yy}(y) \, \d y^2 + G_{\Omega\Omega}(y)\, \d \Omega_6^2
\ee
with a vanishing  $B$-field, which includes \eqq{eqn.D2branes}, \eqq{eqn.HVLifshitz} and \eqq{eqn.AdS}, the DBI action reduces to
\be\label{eqn.D6probeD2}
S_\mt{DBI} = -2\pi^2 \,\td \int \d^3x\, \d y \, e^{-\phi} \sqrt{-\left(G_{tt}G_{yy} + (2\pi\ls^2)^2 \At'^2 \right) G_{xx}^2 G_{\Omega\Omega}^3}  \ ,
\ee
where we have integrated out the three sphere angles, thus producing the $2\pi^2$ factor. Since $\At$ enters only via its derivative,  the solution for this field can be written in terms of a constant of integration, $\nqp$, as
\be
\label{atp}
2\pi \ls^2\, \At' = \sqrt{\frac{\nqp^2 |G_{tt}|G_{yy}}{\nqp^2 + (2\pi^2\,\td\,2\pi\ls^2)^2 e^{-2\phi} G_{xx}^2 G_{\Omega\Omega}^3}} \ .
\ee
This constant is precisely  the string density on the probe, and will play an important role in the rest of the paper. Note that we use a lower-case symbol to distinguish the string density on the probe from that in the background. To work at fixed charge density we  perform a Legendre transform in \eqref{eqn.D6probeD2}, and express the action in terms of $\nqp$ instead of $\At$, obtaining 
\be\label{eqn.legendreprobe}
\widetilde S_\mt{DBI} = -2\pi^2\, \nfp \, \td \int \d^3x\, \d y\, e^{-\phi} \sqrt{-G_{tt} G_{xx}^2 G_{yy} G_{\Omega\Omega}^3}\, \sqrt{1+\frac{e^{2\phi}\, \nqp^2}{(2\pi^2\,\td\,\nfp\,2\pi\ls^2)^2G_{xx}^2 G_{\Omega\Omega}^{3}}} \ , 
\,\,\,\,\,\,\,\,\,\,\,\,\,\,\,\,\,
\ee
where we have included factors of $\nfp$ to account for the possibility of multiple overlapping probe branes. Finally, we compute the stress tensor associated to the D6-brane action. Since the only non-zero component of the gauge field is the temporal one, we focus on the time-time component of the stress tensor. This takes the form 
\be
\widetilde T_{tt} = \frac{-2}{\sqrt{-G}} \,G_{tt}^2\,\frac{\delta \widetilde S_\mt{DBI}}{\delta G_{tt}} = 2\pi^2\,\td\,\nfp\, e^{-\phi}\frac{G_{tt}}{G_{\Omega\Omega}^{3/2}}  \sqrt{1+\frac{e^{2\phi}\, \nqp^2}{(2\pi^2\,\td\,\nfp\,2\pi\ls^2)^2G_{xx}^2 G_{\Omega\Omega}^{3}}}  \,,
\label{probestress}
\ee
where the tilde is a reminder that this is the probe's stress tensor. Our goal is to compare this result to the supergravity stress tensor supporting each of the solutions \eqq{eqn.D2branes}, \eqq{eqn.HVLifshitz} or \eqq{eqn.AdS}, namely to the right-hand side of Einstein's equations
\be
\frac{1}{\kappa^2}E_{\mu\nu} = e^{2\phi} \, T_{\mu\nu} \ ,
\ee
where 
\be
\label{effect}
\frac{1}{2 \kappa^2} = \frac{2\pi}{(2\pi \ell_s)^8 g_s^2}\,, 
\ee
$E_{\mu\nu}$ is the Einstein tensor and the factor of $e^{2\phi}$ appears because we are working in string frame.

Eqns.~\eqq{atp} and \eqq{probestress} would not be modified if we had included the Wess-Zumino (WZ) part of the D6-brane action in our discussion. The reasons are, first,  that for the backgrounds of interest in this section the WZ term does not contain any $A_t$-dependent contributions that could modify our definition \eqq{atp} of $\nqp$ and, second, that the WZ term is topological, i.e.~metric-independent, and therefore it does not contribute to our definition of the probe stress tensor \eqq{probestress}.

We are now ready to determine the crossover scales by comparing the stress tensors. We begin with the D2-brane geometry \eqq{eqn.D2branes}. The probe stress tensor behaves as 
\bse\bal
&e^{2\phi}\,  \widetilde T_{tt}\simeq \frac{\nfp\,\td}{L^5} u^2 & \quad \text{at zero $\nqp$} \ , \label{eqn.probeSEtensornod}\\[1.7mm]
&e^{2\phi} \, \widetilde T_{tt} \simeq  \frac{1}{L^5\,\ls^2} \frac{\nqp}{u} + {\cal O}(u^{5}) & \quad \text{for small $u$ at non-zero $\nqp$}\ , \label{eqn.probeSEtensornodsmallu}\\[2mm]
&e^{2\phi} \, \widetilde T_{tt} \simeq  \frac{\nfp\,\td}{L^5} u^2  + {\cal O}(u^{-4})  & \quad \text{for large $u$ at non-zero $\nqp$} \ , \label{eqn.probeSEtensornodlargeu}
\end{align}\ese
whereas the supergravity stress tensor scales as 
\be\label{eqn.bgSEtensor}
\frac{1}{\kappa^2} E_{tt} \simeq \frac{1}{\kappa^2} \frac{u^3}{L^5} \ .
\ee
Consider first $\nqp=0$. In this case we see that the probe stress tensor is subleading at large $u$, meaning that the flavor branes are a small correction in the UV. The scale at which their stress tensor becomes comparable to the background stress tensor is $u=u_\mt{flavor}$ such that 
\be\label{eqn.probeflavorscale}
\frac{1}{\kappa^2} \frac{u_\mt{flavor}^3}{L^5} \sim \frac{\nfp\, \td}{L^5}u_\mt{flavor}^2  \,.
\ee 
Via \eqq{bigu} this leads precisely to \eqq{uflavor} with $\nf$ replaced by $\nfp$. This is as expected, since this is the scale at which the backreaction of the $\nfp$ flavor probe branes would become important if we were to include it.

Consider now $\nqp \neq 0$. We see that the presence of the strings does not change the UV behavior of the probe stress tensor, meaning that the backreaction of the strings is subleading with respect to that of the flavor branes in this regime. This is consistent with the fall-offs \eqq{chargefalloff} and \eqq{flavorfalloff} mentioned in Sec.~\ref{intro}, and with the fact that the UV geometry is unmodified even in the presence of both charge and flavor. In contrast, in the IR we see that the stress tensor of the brane is dominated by the string density, since the $1/u$ leading term is proportional to $\nqp$ and $\nfp$-independent, and that this would actually dominate over the supergravity stress tensor. This is consistent with the fact that the addition of charge always changes the IR geometry completely, and it drives the RG flow to an HVL solution.  
The  scale at which the crossover takes place is $u=u_\mt{charge}$ such that 
\be\label{eqn.ustprobe}
\frac{1}{\kappa^2} \frac{u_\mt{charge}^3}{L^5} \sim \frac{1}{L^5\,\ls^2} \frac{\nqp}{u_\mt{charge}} \,,
\ee 
which via \eqq{bigu} leads precisely to \eqq{ucharge} with $\nq$ replaced by 
$\nqp$, as expected.
\newline

We now turn to the HVL geometry. The probe stress tensor reads
\bse\bal
&e^{2\phi} \widetilde T_{tt}\simeq \frac{\nfp \,\td\, C_\mt{t}}{\left( \Qst\, L\right)^{3}\, L^3}   \frac{r^{12}}{L^{12}} & \quad \text{at zero $\nqp$} \ , \label{eqn.probeSEtensornodHVLif}\\[2mm]
&e^{2\phi} \widetilde T_{tt} \simeq  \frac{\nqp\, C_\mt{t}}{\ls^2\, (\Qst\,L)^2\, L^6}  \frac{r^{10}}{L^{10}}  + {\cal O}(r^{14}) & \quad \text{for small $r$ at non-zero $\nqp$}\ , \label{eqn.probeSEtensornodsmalluHVLif}\\[3mm]
&e^{2\phi} \widetilde T_{tt} \simeq  \frac{\nfp \,\td\, C_\mt{t}}{\left( \Qst\, L\right)^{3}\, L^3} \frac{r^{12}}{L^{12}} + {\cal O}(r^{8})  & \quad \text{for large $r$ at non-zero $\nqp$} \ , \label{eqn.probeSEtensornodlargeuHVLif}
\end{align}\ese
whereas the supergravity stress tensor is 
\be
\frac{1}{\kappa^2}  E_{tt} \sim \frac{1}{\kappa^2} \frac{ C_\mt{t}}{(\Qst\, L)\, L^2}  \frac{r^{10}}{L^{10}} \ .
\ee
We see that at small $r$ the probe stress tensor is dominated by the string contribution, meaning that the addition of flavor leaves the IR geometry unmodified, in agreement with our previous discussion. Moreover, this string contribution  in the IR scales exactly in the same way as the supergravity stress tensor. This is consistent with the fact that the HVL geometry is sourced by strings, with only subleading contributions from the flavor branes. The radial position  $r_\mt{cross}$ at which the contribution of the flavor to the probe stress tensor becomes comparable to that of the strings, and of course also to that of the background, determines the crossover scale between the string-dominated HVL geometry and  the flavor-dominated AdS$_4$ solution.  This position obeys 
\be 
 \frac{1}{\kappa^2} \frac{C_\mt{t}}{(\Qst\, L)\, L^2} \frac{r_\mt{cross}^{10}}{L^{10}}
\sim 
\frac{\nfp \,\td\, C_\mt{t}}{\left( \Qst\, L\right)^{3}\, L^3}  \frac{r_\mt{cross}^{12}}{L^{12}} \,.
\ee
Using \eqq{eqn.HVLifD2radialchange} and \eqq{bigu} this leads to \eqq{ucross} with $\nf$ replaced by $\nfp$, as expected.
\newline

Finally, we turn to the AdS$_4$ solution. The probe stress tensor is
\bse\bal
&e^{2\phi}\, \widetilde T_{tt}\sim \frac{\nfp \,T_\mt{D6}}{\gs^2\,\ls^5 \nc} R^2 & \quad \text{at zero $\nqp$} \ , \label{eqn.probeSEtensornodAdS}\\[2mm]
&e^{2\phi}\, \widetilde T_{tt} \simeq  \frac{\nqp}{\gs^2\,\ls^8 \, \nf\, \nc} + {\cal O}(R^4) & \quad \text{for small $R$ at non-zero $\nqp$}\ , \label{eqn.probeSEtensornodsmalluAdS}\\[2mm]
&e^{2\phi}\, \widetilde T_{tt} \simeq \frac{\nfp\,T_\mt{D6}}{\gs^2\,\ls^5\, \nc} R^2 + {\cal O}(R^{-2})  & \quad \text{for large $R$ at non-zero $\nqp$} \ , \label{eqn.probeSEtensornodlargeuAds}
\end{align}\ese
whereas the supergravity stress tensor reads
\be\label{eqn.EinsprobeAdS}
\frac{1}{\kappa^2} E_{tt} \sim \frac{1}{\kappa^2} \frac{R^2}{L_\mt{IR}^4} \,.
\ee
If $\nqp=0$ the probe stress tensor scales in the same way as the background stress tensor, consistently with the fact that in this case the background is sourced by a large collection of flavor branes. This remains true at large $R$ even if $\nqp \neq 0$, meaning that the addition of charge leaves the UV of the AdS$_4$ geometry unmodified. In contrast, at small $R$ the strings contribution dominates the probe tensor, and this dominates over the background stress tensor. This confirms that the addition of strings is a relevant deformation from the viewpoint of the fixed point dual to the AdS$_4$ geometry that drives the theory to a new IR geometry. The scale at which the crossover between the two geometries takes place can be determined by comparing  \eqref{eqn.probeSEtensornodsmalluAdS} to \eqref{eqn.EinsprobeAdS}, with the result
\be\label{eqn.Vstar}
R_\mt{cross} \sim \frac{\nqp^{1/2} \, \ell_s^2}{\nf}\ .
\ee
Note that, from the AdS$_4$ prespective, the dependence on $\nqp$ is fixed by dimensional analysis, since this is the only scale at this fixed point. In other words, the theory at this fixed point has lost any memory of the other dimensionful parameter in the gauge theory, the 't Hooft coupling $\lambda$. This means that from the AdS$_4$ viewpoint all values of $\nqp$ are equivalent: they simply set the units in which all other scales such as the crossover scale \eqq{eqn.Vstar} are measured. The dependence on $\lambda$ is recovered when the AdS$_4$ coordinate $R$ is mapped to the IR coordinate $r$ and then to the UV coordinate $u$ via \eqq{eqn.HVLifD2radialchange}. As usual, the relation between $r$ and $R$ is established by matching the $g_{xx}$ components of the dimensionally reduced metrics, which gives 
 \be
\frac{r}{L} = \left(  \frac{27}{8} \frac{\gs \, L_\mt{IR}^2}{L^3} \, 
\frac{\nf^{5/4}}{\nc^{1/4}} \, R \right)^2 \ .
\ee
This relation together with \eqq{eqn.HVLifD2radialchange} and \eqq{bigu} maps $R_\mt{cross}$ to $\uc$. 

\section{Full solutions}\label{sec.setup}
\label{full}
We finally turn to the main result of our paper: the family of supergravity solutions dual to three-dimensional $SU(\nc)$ SYM in the presence of $\nf$ flavors of dynamical quarks and a non-zero quark density $\nq$. The flavor branes and the strings  act as sources for both the RR fields and the $H$-field, and hence they modify their equations of motion. For the RR fields, through the Hodge-duality relations that these fields obey, this also leads to a modification of their Bianchi identities, and therefore to a modification of their very definition in terms of gauge potentials. The full action in the presence of these sources is discussed in Appendix \ref{app_sources}, to which we refer the reader for additional details. 

\subsection{Ansatz}
The full action of the system consists of type IIA supergravity coupled to a set of D6-branes with strings dissolved inside them:
\be\label{eqn.totalaction}
S = S_\mt{IIA} + S_\mt{D6} \,.
\ee
The supergravity action is the sum of the NS and the RR sectors,
\be
S_\mt{IIA} = S_\mt{NS} + S_\mt{RR} \,.
\ee 
Similarly, the D6-action is the sum of the  DBI and the WZ terms:
\be
S_\mt{D6} = S_\mt{DBI} + S_\mt{WZ} \,.
\ee

As we will explain below, the presence of the strings will not change the fact that the metric in the solution is invariant under the full isometry group of the S$^6$, just like when only flavor but no charge is present \cite{Faedo:2015ula}. This means that  the most general metric compatible with the symmetries of our system takes the form \eqq{eqn.metricansatz2}, and also that the solution is naturally equipped with the forms $J$ and $\Omega$ obeying the relations \eqq{eqn.SU3structure} associated to the NK structure of the S$^6$. 

The brane action describes a collection of $\nf$ D6-branes appropriately smeared over the S$^6$ internal directions of the geometry, with a density of strings appropriately smeared within each D6-brane. In the limit in which the number of branes and strings is large, one can approximate their distribution by a continuous function, which will allow us to turn the seven-dimensional D6-brane action into an integral over all spacetime directions. The information about the orientation and the density of branes at each point is encoded in the so-called smearing three-form $\Xi$ that will be determined below. In order to write the smeared DBI part of the action, let us first rewrite the DBI action for a single brane \eqq{single} as
\be
S_\mt{DBI} = - \td \,  \int_{D6}  {\cal P}\left[{\cal K}\right] \,,
\ee
where the seven-form ${\cal K}$ is defined as 
\be
 {\cal K}  = e^{-\phi} \sqrt{-\det\left(G+\mathcal F  \right)} \,\, \d t \wedge \d x^1 \wedge \d x^2 \wedge \d y \wedge \text{Re}\, \Omega \ .
\ee
In terms of ${\cal K}$ and the smearing form, the DBI action for the smeared set of branes takes the form 
\be\label{eqn.sDBI}
S_\mt{DBI} = - T_\mt{D6}  \int  {\cal K} \wedge \Xi \,.
\ee
Similarly, while the WZ part of the action for a single brane is given by 
\be
S_\mt{WZ} = T_\mt{D6}  \int_{D6} {\cal L}_\mt{WZ} \,,
\ee
the smeared action is
\be\label{eqn.sWZ}
S_\mt{WZ} = T_\mt{D6}  \int {\cal L}_\mt{WZ} \wedge \Xi \,.
\ee
The form of ${\cal L}_\mt{WZ}$ can be found in Appendix \ref{app_sources}.

In order to model the presence of the strings, the BI field on the branes takes the same form as in \eqq{at}. Note that the fact that the BI electric field only depends on the radial coordinate $y$ implies that the distribution of the strings respects the full isometry group of the S$^6$.  
As we will see below, the NS field $B$ will vanish in our solution, meaning that 
\be
\label{curlyf}
\mathcal F = 2\pi \ell_s^2 \, \d A = 2\pi \ell_s^2 \, A_t'(y) \, \d y \wedge \d t \,.
\ee
This implies that $\mathcal F \wedge \mathcal F=0$, which leads to a simplification with respect to the general setup discussed in Appendix \ref{app_sources}. Through the WZ part of the branes' action,  $\mathcal F$  modifies the equations of motion for the RR field strengths  $F_8$ and $F_6$ with respect to pure supergravity without sources. Equivalently, it modifies the Bianchi identities for their Hodge duals 
\be
\label{hodge}
F_2=- * F_8 \sac  F_4=*F_6 \,, 
\ee
which in the presence of the D6-branes and the strings read
\bse\bal\label{eqn.F2BIviolation}
\d F_2 & = -2\kappa^2\, T_\mt{D6} \, \Xi \ , \\
\d F_4 & = H\wedge F_2 -2\kappa^2\, T_\mt{D6}\, \mathcal F \wedge \Xi \ . \label{eqn.F4BIviolation}
\end{align}
\ese
The first equation is commonly referred to as the `violation' of the Bianchi identity for $F_2$, and it expresses the simple fact that D6-branes are magnetic sources for $F_2$. This violation was already present in the case of flavor without strings discussed in \cite{Faedo:2015ula}. In contrast, the second term on the right-hand side of the Bianchi identity for $F_4$ was not present for the solutions with strings but without flavor studied in \cite{Faedo:2014ana}. Indeed, this term is only present when the system contains both strings and D6-branes, since it comes from the electric components of $C_5$ sourced by the term 
\be
\int C_5 \wedge \mathcal F
\label{rein}
\ee
contained in the WZ part of the D6-branes' action. Through the duality relations \eqq{hodge} these electric components give rise to the second term on the right-hand side of  $\d F_4$. We thus see that, although it is the string density represented by $\mathcal F$ that sources $C_5$, the presence of the D6-branes is necessary since the coupling between $\mathcal F$ and $C_5$ is supported on their worldvolume.

We are now ready to specify the ansatz for our solution and to integrate the corresponding equations of motion. We begin with the RR forms. We choose to specify directly $F_2$ and $F_6$; their Hodge duals $F_4$ and $F_8$ can be obtained via \eqq{hodge}. Our ansatz for $F_2$ is 
\be\label{eqn.F2form}
F_2 = \Qst \,\d x^1\wedge \d x^2 + \Qf \, J \,.
\ee
This is the most general form compatible with the symmetries of our system, in particular with the $SU(3)$ structure of the S$^6$, except for the fact that we have omitted a term proportional to $\d t \wedge \d y$. This term is allowed by symmetry considerations, but we will show below that it must vanish by virtue of the equation of motion for the $B$-field.  As we will see, the constant  $\Qst$ is related to the string density \cite{Chen:2009kx}. Similarly, $\Qf$ is related to the distribution of D6-branes along the radial direction. If some of the quark flavors in the gauge theory are massive, then $\Qf$ becomes a  function of the radial direction, $\Qf(y)$. Here we will restrict ourselves to the case in which all quarks are massless, which translates into the fact that $\Qf$ is a constant related to the number of D6-branes through \eqq{eqn.Qfdef}. Computing $\d F_2$ with \eqq{eqn.F2form} and \eqq{eqn.SU3structure} and comparing to \eqref{eqn.F2BIviolation} we deduce that the smearing  form is
\be
\label{xixi}
\Xi = -\frac{3\,\Qf}{2\kappa^2\, T_\mt{D6}}\, \text{Im}\,\Omega \ .
\ee

In order to specify $F_6$, we first note from \eqq{curlyf} and \eqq{xixi} that the  coupling in \eqq{rein} implies that $C_5$ must contain a term of the following form:
\be\label{eqn.C5potential}
C_5 \supset \bfunc(y) \, \d x^1 \wedge \d x^2 \wedge \text{Re}\,\Omega \,.
\ee
Including also the term necessary to account for the presence of D2-branes in the solution, as in \eqq{eqn.D2sRR}, we see that $F_6$ must take the form
\be\label{eqn.F6form}
F_6 = \frac{\Qc}{6} J \wedge J \wedge J \,+ \, \bfunc'\,  \d y \wedge \d x^1 \wedge \d x^2 \wedge \text{Re}\,\Omega \,+ \, 2 \bfunc \,\d x^1 \wedge \d x^2 \wedge J \wedge J \ ,
\ee
where $'$ denotes differentiation with respect to $y$ and we have used \eqq{eqn.SU3structure} to write the volume form on the six-sphere in terms of $J$. The constant $\Qc$ is related to the number of D2-branes through \eqq{eqn.Ncquantization}. The equation of motion for the function $\bfunc(y)$ is obtained by Hodge-dualizing $F_6$ to $F_4$ with the metric \eqq{eqn.metricansatz2}, which yields
\begin{eqnarray}
\label{f4}
F_4 &=& \Qc \frac{\sqrt{-G_ {tt} \, G_{yy}} \, G_{xx}}{G_{\Omega\Omega}^3}\, \d t \wedge \d x^1 \wedge \d x^2 \wedge \d y   \nonumber \\[2mm]
&& +  4 \, \cB \, \frac{ \sqrt{-G_{tt} \, G_{yy}} }{ G_{xx}\, G_{\Omega\Omega} } \d t \wedge \d y \wedge J  \nonumber \\[3mm]
&&+  \cB'  \frac{ \sqrt{-G_{tt}} }{ \sqrt{ G_{yy}} \, G_{xx} } \d t \wedge \text{Im} \Omega \,,
\end{eqnarray}
 and then substituting into the Bianchi identity \eqq{eqn.F4BIviolation}. The result is 
\be\label{eqn.bfuncequation}
\partial_y \left( \sqrt{\frac{-G_{tt}}{G_{yy}}} \, \frac{\bfunc'}{G_{xx}} \right) - 12 \frac{\sqrt{-G_{tt}\, G_{yy}}}{G_{xx}\,G_{\Omega\Omega}}\, \bfunc - 3\, \Qf \, 2\pi\ls^2\, \At' = 0 \ .
\ee
The physical meaning of the `magnetic potential' $\cB$ is more easily understood by rewriting it in terms of an electric, dual potential. For this purpose, we write the following ansatz for the three-form RR potential:
\be
C_3 = \widetilde C_3+\cfunc(y) \,\d t \wedge J \,,
\ee
where $\widetilde C_3$ is the usual piece associated to the presence of D2-branes that obeys
\be
\d \widetilde C_3 =  \Qc \frac{\sqrt{-G_ {tt} \, G_{yy}} \, G_{xx}}{G_{\Omega\Omega}^3}\, \d t \wedge \d x^1 \wedge \d x^2 \wedge \d y \,.
\ee
Substituting $\d C_3$ in the definition \eqq{eqn.newforms} of the modified field strength $F_4$ and comparing to \eqq{f4} we see that $\cfunc(y)$ and its first derivative must satisfy the equations
\be
\cfunc = - \frac{1}{3} \sqrt{\frac{-G_{tt}}{G_{yy}}} \frac{\bfunc'}{G_{xx}} \ , \qquad \cfunc'+\Qf\, 2\pi\ls^2\,\At' = -4 \frac{\sqrt{-G_{tt}\,G_{yy}}}{G_{xx}\,G_{\Omega\Omega}} \bfunc \,.
\ee
These equations allow us to eliminate $\bfunc$ in favor of $\cfunc$ and vice versa. The term in $C_3$ proportional to $\cfunc$ can now be interpreted as associated to D2-branes wrapped on the cycle threaded by $J$. Upon reduction along the compact directions, this  gives rise to a massive vector field with positive mass squared.\footnote{An analogous field was present in the four-dimensional case of \cite{Bigazzi:2011it} and its dual interpretation around the AdS$_5$ region was given in \cite{Cotrone:2012um}.} Therefore the field $\cfunc (y)$ is dual to (the time component of) an irrelevant vector operator constructed  out of adjoint fields in the gauge theory. This should be contrasted with the vector field $A$ on the D6-branes, which is dual to a conserved, and therefore marginal, current operator in the gauge theory constructed out of the microscopic fields in the fundamental representation of the gauge group. Below we will have more to say  about the interplay between  these two vector fields. 

We now turn to the equation of motion for the $B$-field. This may be written as
\be
\label{eqn.eomAt}
\d \left( e^{-2\phi} * H \right) = \frac{\delta S}{\delta B}
\,,
\ee
where the right-hand side includes all variations with respect to $B$ but not $\d B$, and is given by
\bal
\label{eqn.eomAt2}
\frac{\delta S}{\delta B} & = F_2 \wedge *F_4 \,+ \,\frac{1}{2} F_4 \wedge F_4  \\
& \quad + 2\kappa^2\, T_6\, e^{-\phi}\sqrt{G_{xx}^2 G_{\Omega\Omega}^3} \frac{2\pi\ls^2\, \At'}{\sqrt{- \left( G_{tt} G_{yy} + (2\pi \ls^2 \At')^2 \right)} } \d x^1 \wedge \d x^2  \wedge \text{Re}\,\Omega \wedge \Xi \ . \nonumber
\end{align}
The second line in this equation  is the DBI contribution. 
We will solve this equation with $B=H=0$, which immediately implies that the term proportional to $\d t \wedge \d y$ allowed by symmetry considerations in \eqq{eqn.F2form} must vanish.  At first sight it may seem surprising that the presence of strings does not automatically lead to a non-zero 
$H$, but the non-linearities of supergravity imply that the $H$ sourced by the strings can be exactly cancelled by the $H$  sourced by the  products of RR forms in \eqq{eqn.eomAt2}. Requiring this cancellation fixes the BI field on the D6-branes to 
\be\label{eqn.Atequation}
2\pi\ls^2 \, \At' =  \sqrt{-G_{tt} G_{yy}}   \frac{  e^{\phi} \, G_{\Omega\Omega}^{-3/2}  \left(\Qc\, \Qst + 12 \, \Qf\,  \bfunc \right) }{\sqrt{ \left( 12 \, \Qf\, G_{xx} \right)^2 + e^{2\phi} \, G_{\Omega\Omega}^{-3} \left( \Qc\, \Qst + 12 \, \Qf\,  \bfunc\right)^2}} \ .
\ee

To close the circle we note that  \eqref{eqn.Atequation} is automatically a solution of the equation of motion for the BI  field obtained by varying the full action with respect to $A$. The reason is that $\d A$ always appears together with $B$ in the gauge-invariant combination $\d A + B$. This means that   
\be 
\frac{\delta S}{\delta \d A} = \frac{\delta S}{\delta B}  
\ee
and therefore that  the equation of motion for $A$, 
\be 
\d \, \frac{\delta S}{\delta \d A} = 0 \,,
\ee
is automatically implied by the exterior derivative of \eqq{eqn.eomAt}. In fact, substituting  \eqq{eqn.Atequation} in \eqq{eqn.eomAt2} gives the explicit result for the so-called electric displacement 
\be\label{eqn.electricdisplacement}
\frac{\delta S}{\delta \d A}   = 2\pi \ls^2\,\frac{\Qc \, \Qst}{2\kappa^2} \,
\d x^1 \wedge \d x^2 \wedge \omega_6 \,.
\ee
The string density in the $x^1-x^2$ directions is obtained by integrating this expression over the six-sphere, 
\be
\nq \, \d x^1 \wedge \d x^2 = \int_{\text{S}^6} \frac{\delta S}{\delta \d A}   \,,
\ee
which finally yields the relation between the string density and $\Qst$:
\be\label{eqn.Qstring}
\Qst  = \frac{2\kappa^2}{\Qc} \frac{\nq}{2\pi\ls^2\,V_6} = 2\pi \ls^2\, \lambda\, \frac{\nq}{  \nc^2} \ .
\ee

Note that $\nq$ is the total string density in the system, whereas the string density per D6-brane is $q_\mt{st} = \nq / \nf$. Thus one may formally consider a limit in which the number of D6-branes goes to zero and the string density per D6-brane diverges in such a way that the total string density is kept fixed: 
\be
\nf \to 0 \sac q_\mt{st} \to \infty \sac \nq \,\,\, \mbox{fixed} \,. 
\ee
In this limit $\bfunc=0$ and the model presented in this section reduces to the one in  \cite{Faedo:2014ana}, which describes the flow along the vertical line on the left-hand side of the triangle in Fig.~\ref{fig.RGflows}.  In particular, the DBI term \eqref{eqn.sDBI} reduces to a smeared Nambu-Goto action for a uniform distribution of strings stretching along the radial direction.  
In the opposite limit in which we set $\Qst=\bfunc=0$ the model of this section reduces to that of \cite{Faedo:2015ula}, where the supersymmetric flow with flavor but no charge represented by the upper diagonal edge of the triangle in  Fig.~\ref{fig.RGflows} was constructed.

\subsection{Scalings}
\label{scalings}
In principle, the physics in our system depends on the four parameters $\lambda, \nc, \nf$ and $\nq$. However, through appropriate rescalings we will now see that the physics depends non-trivially only on a dimensionless combination of these parameters. In order to show this, it is useful to begin by recalling the length dimensions of the dimensionful variables in our set-up: 
\be
[y]\sim \ell \ , \quad [\Qc] \sim \ell^5 \ , \quad [\Qf] \sim \ell \ , \quad[\Qst] \sim \ell^{-1} \ , \quad[\bfunc] \sim \ell^3 \ , \quad[G_{\Omega\Omega}] \sim \ell^2 \ , \quad [\At] \sim \ell^{-1} \ .
\ee
The metric components $G_{tt}, G_{xx}, G_{yy}$ and  the dilaton are dimensionless.  

In order to work  with dimensionless quantities we need to choose a specific unit of length. Any combination of $\Qc$, $\Qf$ and $\Qst$ with length dimensions is a valid choice, and we find it convenient to simply use $\Qf$ (which implies that taking the $\nf\to0$ limit in the dimensionless variables is not straightforward). We thus write the radial coordinate and the dimensionful functions in our ansatz in terms of the following dimensionless counterparts:
\be
\label{rescresc}
y = \Qf \, \overline{y} \ , \qquad G_{\Omega\Omega} = \Qf^2\, \widetilde G_{\Omega\Omega} \ , \qquad \bfunc = \Qf^3 \, \overline \bfunc \ .
\ee
The reason for using a tilde on $G_{\Omega\Omega}$ instead of an overline  will become clear shortly.
Upon implementing this transformation in the radicand in 
Eqn.~\eqref{eqn.Atequation}, we observe  that the following dimensionless quantity measures the effect of the charge density:
\be\label{eqn.dimensionlessfactor}
\ratio \equiv \frac{\Qc\, \Qst}{12\, \Qf^4} = \frac{\pi^2\, V_2^4}{3\, V_6} \,\frac{1}{\lambda^{2}}\, \left( \frac{\nc}{\nf} \right)^4 \frac{\nq}{\nc^2} \ ,
\ee
where we recall that $V_2$ and $V_6$ are dimensionless volumes. Note that all the dependence on the string theory parameters $\ell_s, g_s$ has cancelled, indicating that $\rho$ is directly a gauge theory parameter. We will now show that the physics depends non-trivially only on this parameter.  

In order to do so, we rescale some of our functions with the following powers of the second dimensionless quantity $\Qc/\Qf^5$:
\bse
\label{redef}
\bal
G_{tt} & = \left(\frac{\Qc}{\Qf^5}\right)^{-1/2} \, \overline G_{tt} \ , \qquad G_{xx} =  \left(\frac{\Qc}{\Qf^5}\right)^{-1/2} \, \overline G_{xx} \ , \\
G_{yy} & =  \left(\frac{\Qc}{\Qf^5}\right)^{1/2} \, \overline G_{yy} \ , \,\quad\quad \widetilde G_{\Omega\Omega} =  \left(\frac{\Qc}{\Qf^5}\right)^{1/2} \, \overline G_{\Omega\Omega} \ , \\
e^\phi & = \left(\frac{\Qc}{\Qf^5}\right)^{1/4} \,e^{\overline\phi}  \ .
\end{align}\ese
Note that $\widetilde G_{\Omega\Omega}$ gets rescaled to its final counterpart $\overline G_{\Omega\Omega}$. Through Eqn.~\eqref{eqn.Atequation}, the rescalings above imply that $A_t'$ does not get rescaled i.e.~that 
\be
A_t' = \overline{A}_t'
\label{nores}
\ee
with 
\be\label{eqn.AtequationBis}
2\pi\ls^2 \,  \overline{A}_t'=  \sqrt{- \overline G_{tt} \overline G_{yy}}   
\frac{e^{\overline \phi} \,\overline G_{\Omega\Omega}^{-3/2}  
\left( \rho + \overline \bfunc \right) }{\sqrt{ \overline G_{xx}^2 + 
e^{2 \overline \phi} \, \overline G_{\Omega\Omega}^{-3} 
\left( \rho +  \overline \bfunc\right)^2}} \ .
\ee

When the original functions and radial coordinate are replaced by their overlined  counterparts, all the dependence on $\Qf$ and $\Qc$ in the action \eqref{eqn.totalaction} cancels out, leaving behind only a dependence on $\rho$ except for an overall factor of $\Qf^5$ in front of the action. 
The effective gravitational coupling $\Qf^5/2\kappa^2$ thus  has dimension $\ell^{-3}$, since the only dimensionful coordinates to integrate over after the rescaling are the three Minkowski coordinates. This result means that the equations of motion in terms of the rescaled variables depend only on $\rho$, as we wanted to show. In fact, by performing a further rescaling of the form 
\be
\overline G_{xx} \to \ratio\, \overline G_{xx} \sac
\overline \bfunc \to \ratio\, \overline \bfunc \,,
\ee
we could eliminate the dependence on $\rho$ from the equations of motion, but only at the expense of introducing $\rho$ dependence in the boundary conditions. We will therefore not perform this rescaling.

We conclude that only the parameter $\rho$ distinguishes  inequivalent RG flows  (i.e.~flows that cannot be mapped to one another by simple rescalings) in the space of theories parametrized by $\lambda, \nc, \nf$ and $\nq$. This parameter can be naturally understood as the ratio between the typical energy scales associated to charge and flavor, as anticipated in Eqn.~\eqq{rhoratio}.
Unless stated otherwise, in the rest of the paper we will work with the dimensionless (overlined) variables, and we  will omit the overlines for simplicity. The full set of equations of motion written in terms of these variables can be found in Appendix \ref{eomG}.

\subsection{Numerical integration}
\label{numint}
In this section we will provide an overview of how we integrated the equations of motion numerically, relegating some technical details and long equations to Appendix \ref{num}. We will begin with equations in which $\lambda, \nc, \nf$ and $\nq$ have not been eliminated yet, and later we will explain how to exploit  the fact that the physics only depends on $\rho$ to construct the solutions. 
 
To obtain a numerical solution, it is convenient to parametrize the string-frame metric and the dilaton as
\begin{align}\label{eq:ansatz}
&\d s^2  = h^{-\frac{1}{2}} \left( - f_1 \d t^2 + \d x_2^2 \right) +h^{\frac{1}{2}} e^{2\chi} \left( \frac{\d y^2}{f_2} + y^2 \d s_6^2 \right) \,,\nonumber\\
&e^{\phi} =h^\frac{1}{4} e^{3\chi} g \,,
\end{align}
where $f_1,f_2, h,\chi$ and $g$ depend only on the radial coordinate $y$. This choice of  parametrization is motivated by the fact that it generalizes that of the chargeless case \cite{Faedo:2015ula} by allowing for the breaking of Lorentz invariance expected in the presence of a charge density. Using diffeomorphism invariance, we fix the radial coordinate by setting 
\be
h=\frac{\Qc}{5 y^5} \,, 
\ee
as in the D2-brane solution. Fixing the form of $h$  allows us to solve algebraically for the function $f_2$, whose expression is given in \eqq{f2fun}. Moreover, the gauge field $A_t$ on the D6-branes can be solved for in terms of the other functions, as given in \eqq{att}, and eliminated from the remaining equations. 

We are thus left with a set \eqq{four} of four coupled, second-order differential equations for $f_1, \chi, g$ and $\mathcal{B}$ that we must solve numerically, subject to the boundary conditions that they interpolate between the D2-brane solution in the UV and the HVL solution in the IR. The asymptotic solution in the UV, 
Eqn.~\eqq{eqn.UVexp}, depends on four undetermined constants associated to the VEVs of the following four operators: the stress tensor, $\mbox{Tr} F^2$, $\mbox{Tr} F^4$, and the operator dual to $\bfunc$.  Similarly, the asymptotic solution in the IR \eqq{eqn.IRexp} depends on another four undetermined constants. The match between the number of undetermined constants is a necessary consistency check for the existence of solutions connecting these two asymptotic behaviors. 

A further consistency check comes from requiring the existence of flows that come very close to the AdS$_4$ fixed point in Fig.~\ref{fig.RGflows}. Any such flow must be driven from the fixed point to the UV by modes that grow (or at least stay constant) towards the UV, and from the fixed point to the IR by modes that grow (or at least stay constant) towards the IR. 
This immediately implies that the  fields that are turned on around the fixed point cannot be dual to relevant operators. Indeed, a field dual to an operator of dimension $\Delta$ behaves as a combination of $\varrho^{d-\Delta}$ (dual to a source) and  $\varrho^{-\Delta}$ (dual to a VEV), where $\varrho$ is the usual Fefferman-Graham  coordinate and the AdS boundary is at $\varrho \to \infty$. For a relevant operator both modes decay towards the UV and grow towards the IR,  so the presence of such an operator would preclude the necessary matching. In contrast, irrelevant or marginal operators are allowed.  In Appendix \ref{spectrum} we compute the spectrum of fluctuations around the AdS$_4$ fixed point and verify that no relevant operator is turned on. 

We  choose to solve the system of four coupled second-order ODEs using a relaxation method. We begin by compactifying the radial coordinate using the transformation 
\be\label{eq.compacty}
y=\frac{1+Y}{1-Y} \,,
\ee
so that our functions are defined on the domain $Y \,\in \,[-1,1]$. For numerical convenience, we redefine the unknown functions by removing their IR scaling behavior as follows:
\bse
\begin{align}
& \mathcal{B}=\left(\frac{1+Y}{2}\right)^{10/3}\widetilde{\mathcal{B}}\,,\\
&f_1=\left(\frac{1+Y}{2}\right)^{20/3}\tilde f_1\,,\\
&g=\left(\frac{1+Y}{2}\right)^{4/3} \tilde g\,,\\
&e^\chi=\left(\frac{1+Y}{2}\right)^{2/3}{e^{\tilde \chi}}\,.
\end{align}
\ese
The advantage of this redefinition lies in the fact that it simplifies the implementation of the boundary conditions. To be more precise, the boundary conditions that we need to impose are 
\be
\{\widetilde{\cal B},\tilde f_1,\tilde g,{e^{\tilde \chi}}\}=\{0,1,1,1\} 
\ee
in the UV ($Y=1$) and 
\be
\{\widetilde{\cal B},\tilde f_1',\tilde g,{e^{\tilde \chi}}\}=\{-\frac{3\cdot 5^{2/3} \Qf}{(14 \Qc \Qst)^{1/3}},0,\frac{5^{1/6} 14^{2/3}}{ \sqrt{11}(\Qc \Qst)^{1/3}}, \frac{10^{1/3}}{(7 \Qc \Qst)^{1/6}}\}  
\ee
in the IR ($Y=-1$). Note that we are fixing the derivative of $f_1$ to zero in the IR. The value of $f_1$ itself in the IR is determined by the flow and is given by the constant $C_\mt{t}$ in Eqn.~\eqq{eqn.IRexp}. We will see below that this constant plays an important role in the low-temperature thermodynamics of the system. 
\begin{figure}[t!!!]
\centering
{\includegraphics[width=11cm]{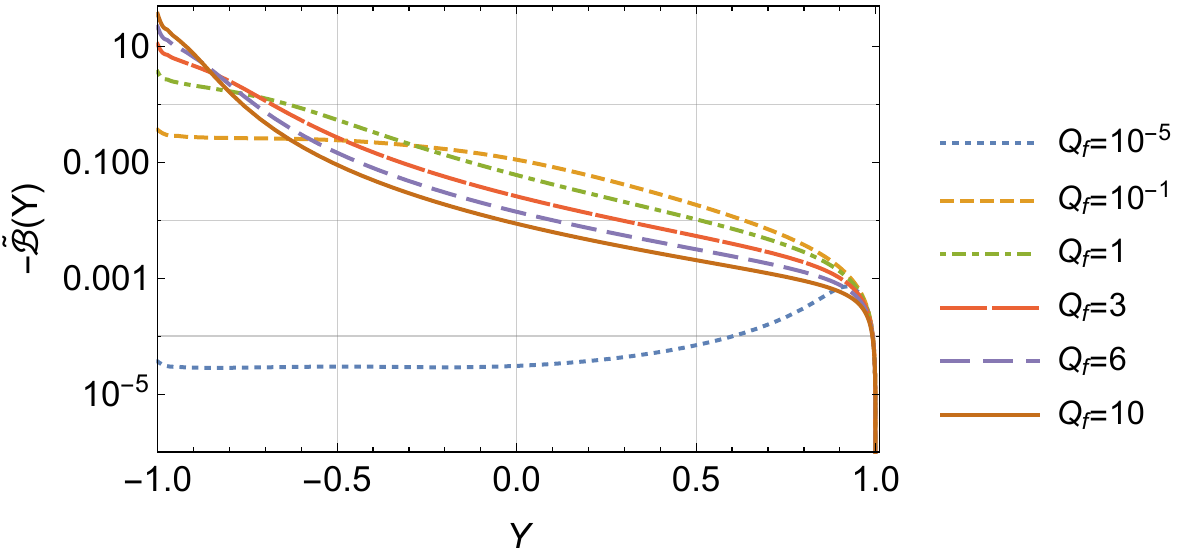}}
{\includegraphics[width=11cm]{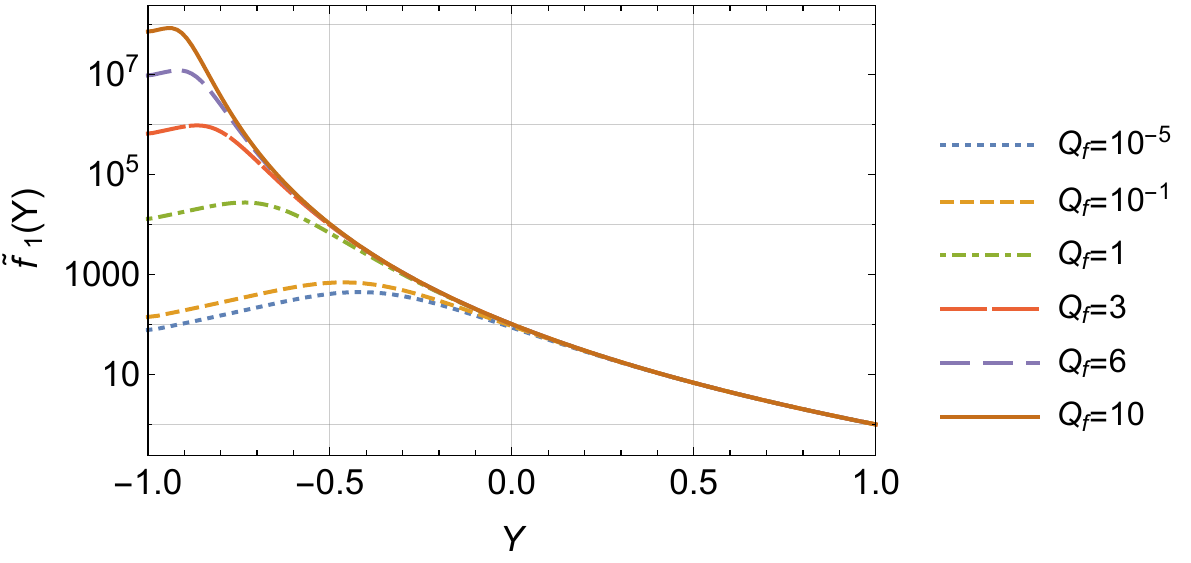}}
{\includegraphics[width=11cm]{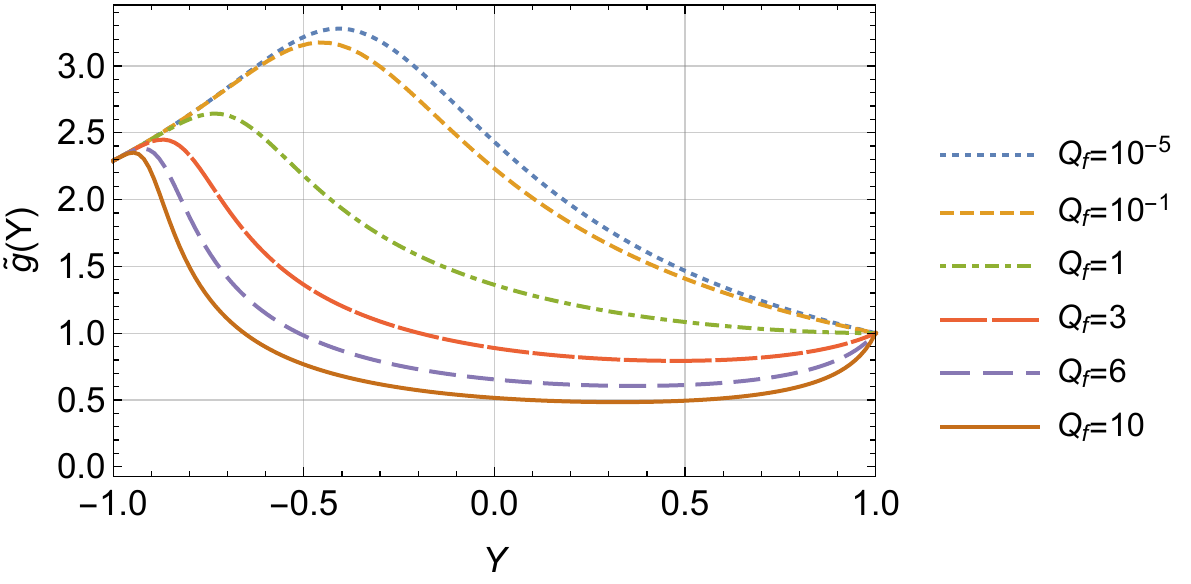}}
{\includegraphics[width=11cm]{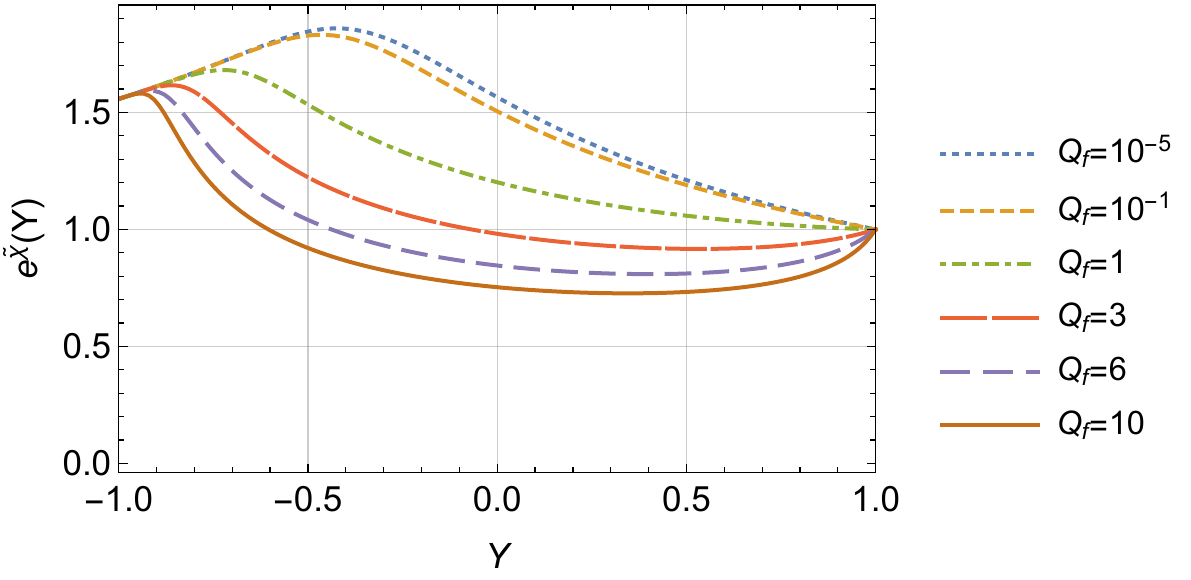}}
\caption{\small Plots of the functions $\widetilde{\cal B}(Y),\tilde f_1(Y),\tilde g(Y), {e^{\tilde \chi}}(Y)$ for several flows with $\rho=1/12 \Qf^4$.}\label{fig:functions}
\end{figure}

We proceed by evenly discretising the coordinate $Y$ to form a lattice and we approximate the derivatives of our functions at each lattice point using a fourth-order finite-difference scheme.\footnote{ We choose to work with finite differences  over pseudospectral methods because of the logarithms that appear in the UV expansion \eqref{eqn.UVexp}.}  The problem is then reduced to solving a set of non-linear algebraic equations for the values of each function at each point on the grid. This is done using the Newton-Raphson method: one starts with an initial guess for the unknown functions at each lattice point, which presumably does not solve the differential equations, and then iteratively improves the guess in order to obtain functions that solve the equations to the desired accuracy.

\begin{figure}[t!!!]
\centering
{\includegraphics[width=12cm]{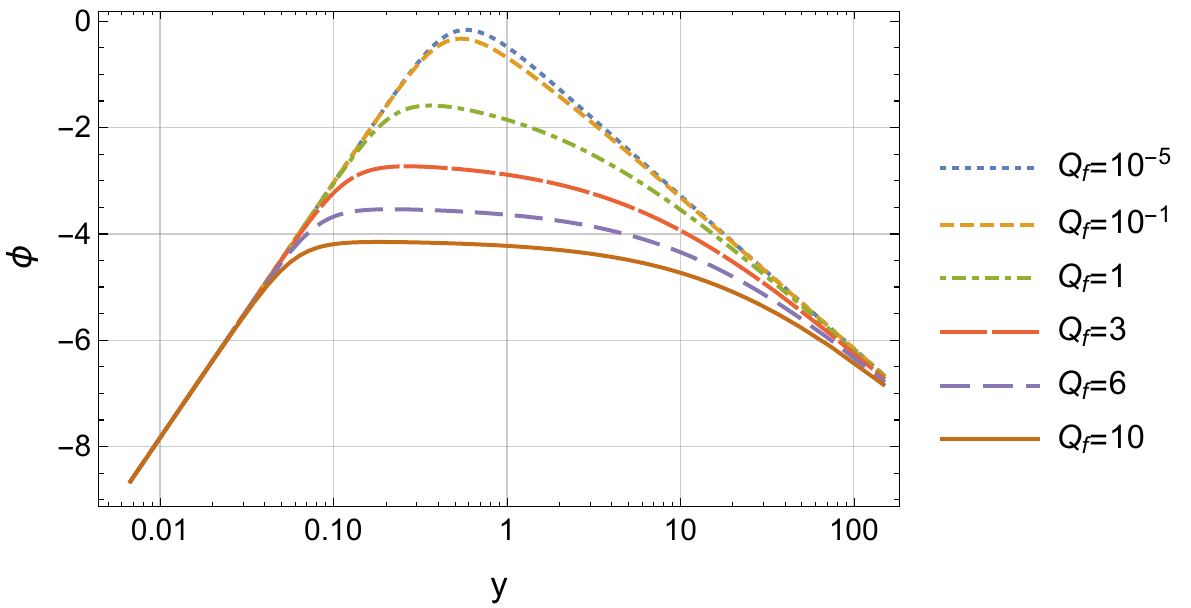}}\quad {\includegraphics[width=13cm]{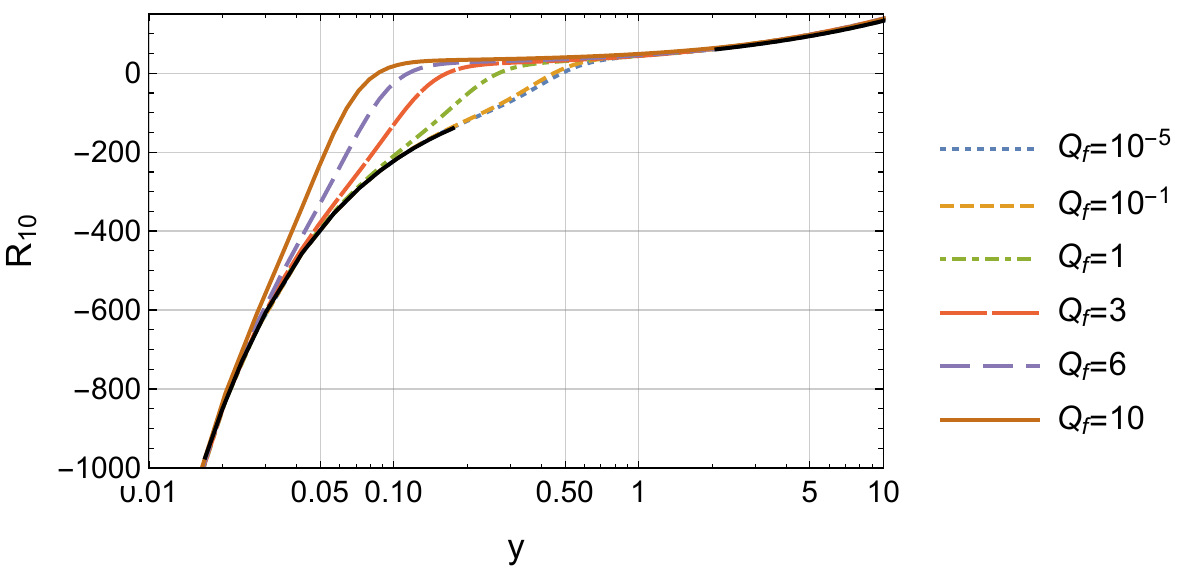}}
\caption{\small Plots of the dilaton $\phi(y)$ and the ten-dimensional Ricci scalar $R_\mt{10}(y)$ in the string frame for several  solutions with $\rho=1/12 \Qf^4$. In the lower panel, the continuous  black curves correspond to the HVL solution (at small $y$) and to the D2-brane solution (at large $y$),  respectively.}
\label{fig:DilatonRicci}
\end{figure}

Given the scaling \eqref{redef}, we can restrict our attention  without loss of generality to solutions with $\Qc=\Qst=1$  and $0<\Qf<\infty$. Each of these solutions then provides a representative of  a flow with a value of $\rho$ given by $\rho=1/12 \Qf^4$, from which the most general solution with arbitrary values of all the parameters can be obtained through the rescalings discussed in Sec.~\ref{scalings}.
The strategy for finding these solutions is to start by picking 
$\Qf=0$ ($\bfunc=0$), describing the flow \cite{Faedo:2014ana} between the D2-brane in the UV and the HVL in the IR, and then slightly deform this solution by introducing a small amount of flavor. This is feasible because in both cases the end points of the integration are the same. Once a solution with a small but non-vanishing $\Qf$ is obtained, one can slowly increase the flavor  using the previous solution as an initial guess for the next one. The profiles of the functions for solutions with various values of $\Qf$ are shown in Fig.~\ref{fig:functions}, while in Fig.~\ref{fig:DilatonRicci} we plot the dilaton and the ten-dimensional Ricci scalar in the string frame.

\section{Quasi-conformal dynamics and Wilson loops}
\label{quasi}
In Fig.~\ref{fig:DilatonRicci} we see that the dilaton profile develops a large plateau in the radial direction (note the logarithmic scale on the horizontal axis), in which it becomes approximately constant, for flows with large values of $\Qf$, or equivalently with small values of $\rho$. Given that radial position in the bulk maps to energy scale in the gauge theory,  this plateau should correspond to the `walking' energy region in the gauge theory in which the dynamics is quasi-conformal, as expected for small-$\rho$ flows that come close to the AdS fixed point in Fig.~\ref{fig.RGflows}. However, one must recall that the profile of the dilaton as a function of the radial coordinate is not a gauge-invariant quantity, since it can be changed by a reparametrization of the radial coordinate. 

\begin{figure}[t!!!]
\begin{center}
\includegraphics[width=0.80\textwidth]{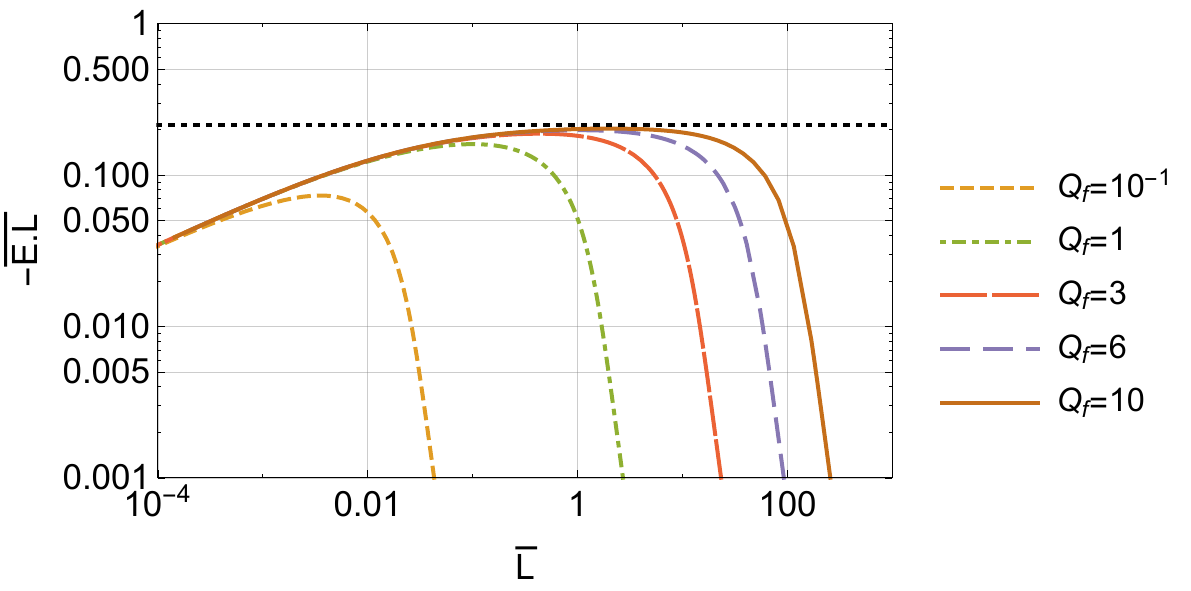}
\caption{\small \label{fig:WL}
Dimensionless potential energy  for an external quark-antiquark pair times the separation distance, $\overline E \cdot \overline L$, as a function of their dimensionless separation, $\overline L$, for various geometries with \mbox{$\rho=1/12 \Qf^4$.} The dotted horizontal line corresponds to the value \eqq{ELads} at the AdS$_4$ fixed point.}
\end{center}
\end{figure}

In order to establish the existence of a walking region in the gauge theory, one must therefore compute a gauge-invariant observable that exhibits quasi-conformal dynamics directly as a function of a gauge theory energy or length scale. For this purpose we have computed the quark-antiquark potential between two external sources as a function of their separation, $L$. As usual, this can be extracted from the expectation value of a Wilson loop, which can be computed on the gravity side from the length of a fundamental string hanging from the quark and the antiquark \cite{Maldacena:1998im,Rey:1998ik}. The details of the calculation follow those in the original papers above, so we have relegated them to Appendix \ref{wilson}. Here we will simply describe the result, which is summarized by Fig.~\ref{fig:WL}.
To obtain this figure it is convenient to add to the rescalings   \eqref{rescresc} and  \eqref{redef} a further rescaling of the $x$-coordinate given by 
\be
x \to \Qf \left( \frac{\Qc}{\Qf^5} \right)^{1/2} \overline x \,.
\label{x_scaling}
\ee
Under these conditions the energy of the quark-antiquark pair and their separation scale as
\be
E = \frac{\Qf}{2\pi \ell_s^2} \overline E(\rho) \sim \lambda \frac{\nf}{\nc} \overline E(\rho) \ , \qquad L = Q_f \left( \frac{\Qc}{\Qf^5} \right)^{1/2} \overline L(\rho) \sim \frac{1}{\lambda} \left( \frac{\nc}{\nf} \right)^{3/2} \overline L(\rho)\ .
\ee
We thus see that the scalings \eqref{rescresc}, \eqref{redef} and \eqq{x_scaling} allow us to reduce the problem of computing the quark-antiquark potential in a  four-parameter flow with generic values of $\nc, \nf, \nq$ and $\lambda$ to a calculation in a  flow characterized by a single parameter $\rho$. 

Fig.~\ref{fig:WL} shows that for large-$\rho$ flows (small $\Qf$) the dependence of the quark-antiquark potential transitions directly from $\overline E \sim \overline L^{-2/3}$ in the UV (small $\overline L$) to $\overline E \sim \overline L^{-6}$ in the IR (large $\overline L$). The former behavior is that of the D2-brane geometry, whereas the latter is what one obtains for the HVL solution. In contrast, for small-$\rho$ flows (large $\Qf$) we see that a large intermediate plateau appears (note the logarithmic scale on the horizontal axis). In this region the potential behaves as 
$\overline E \cdot \overline L^{-1} \sim \mbox{constant}$,  as expected in the quasi-conformal region near a fixed point. The precise value of this constant can be computed using Eqn.~\eqq{E10} and the rescaled version of \eqq{LLL} and is given by
\be
\label{ELads}
- \overline{E} \cdot \overline{L} = \frac{ 16\pi }{ 27 } \frac{ \Gamma\left( \frac{3}{4} \right)^2 }{ \Gamma\left( \frac{1}{4} \right)^2 } \simeq 0.213 \,.
\ee
This result, represented by the dotted horizontal line in Fig.~\ref{fig:WL},  agrees perfectly with the location of the plateau.

\section{Thermodynamics}
\label{thermo}
In this section we will study several thermodynamic properties of our system. First we will compute the chemical potential, the equation of state and the speed of sound at zero temperature.  Then we will consider a small temperature and determine the entropy density.

\subsection{Chemical potential and equation of state}
\label{chemical}
The chemical potential $\mu$, conjugate to the quark density, is given by the work that is necessary to take a unit of charge from the IR to the UV, namely by
\begin{equation}
\mu=\int_{y=0}^{y=\infty} A_t'(y) \, \d y \,.
\end{equation}
Using  \eqq{rescresc} and \eqq{eqn.AtequationBis} this can be written in terms of the rescaled variables as
\be
\mu = \frac{\Qf}{2\pi \ell_s^2} \, \overline \mu (\rho) = 
\frac{\lambda}{V_2}  \frac{\nf}{\nc} \, \overline \mu (\rho)
\ee
with 
\be \label{eq.mubar}
\overline \mu (\rho) = \int_{\overline y=0}^{\overline y=\infty} \d \overline y
\sqrt{- \overline G_{tt} \overline G_{yy}}   
\frac{e^{\overline \phi} \,\overline G_{\Omega\Omega}^{-3/2}  
\left( \rho + \overline \bfunc \right) }{\sqrt{ \overline G_{xx}^2 + 
e^{2 \overline \phi} \, \overline G_{\Omega\Omega}^{-3} 
\left( \rho +  \overline \bfunc\right)^2}} \,.
\ee
\begin{figure}[t]
\begin{center}
 \includegraphics[width=0.6\textwidth]{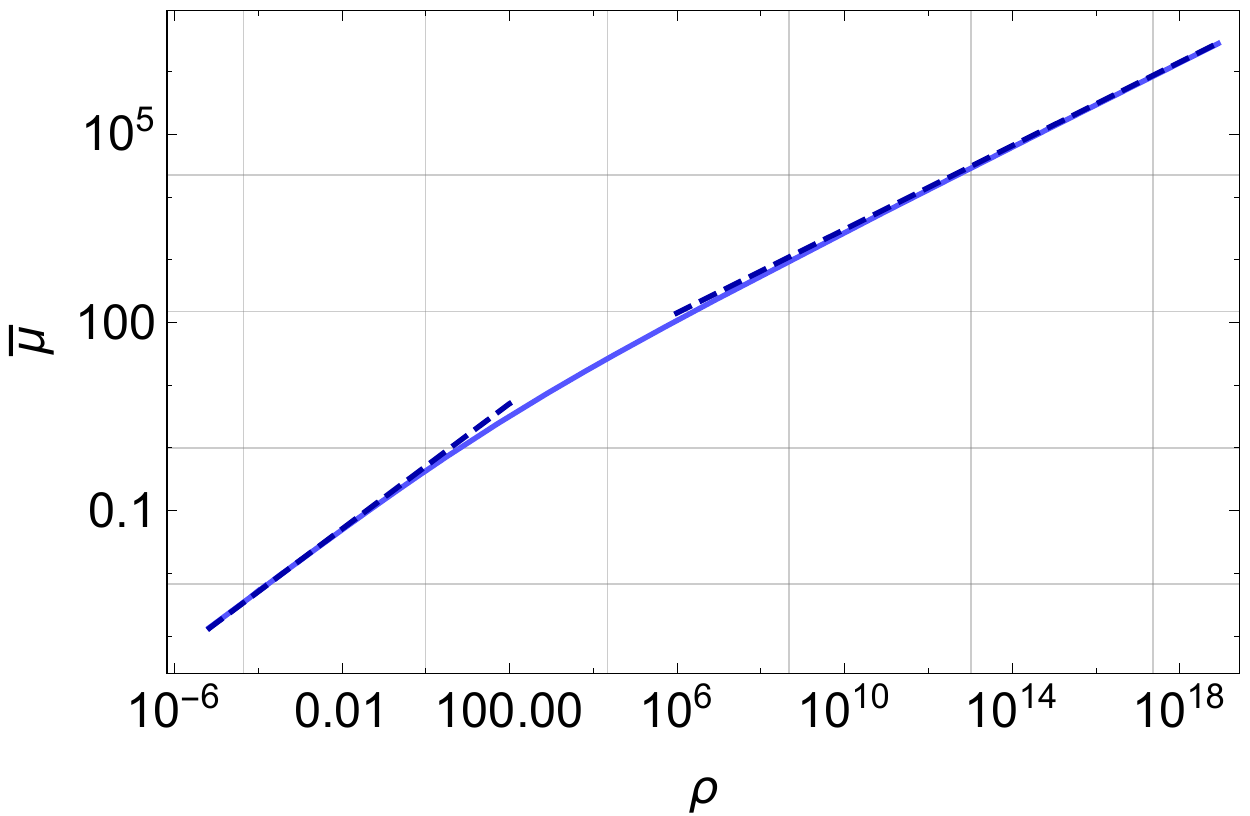}
\caption{ \label{fig:IRdof_chemicalPot}\small
Dimensionless chemical potential \eqq{eq.mubar} as a function of $\rho$. We show dashed lines with behavior $\overline \mu=0.5\, \rho^{1/2}$ at small $\rho$ and $\overline \mu \sim1.4 \,\rho^{1/3}$ at large $\rho$.}
\end{center}
\end{figure}
We see that we have again reduced the calculation for generic values of the several parameters of the theory to a calculation of a function of a single parameter $\rho$. 

The result of evaluating the integral \eqq{eq.mubar} numerically is  plotted in 
Fig.~\ref{fig:IRdof_chemicalPot}.
We see that at small values of $\rho$ the chemical potential scales as
 \be
 \label{eq.musmallrho}
 \overline \mu \sim \rho^{1/2} \sac \mbox{i.e.} \qquad 
\mu\sim  \frac{\nq^{1/2}}{\nf} , 
 \ee
whereas at large values of $\rho$ we have
\be
\label{eq.mulargerho}
\overline \mu \sim \rho^{1/3} \sac \mbox{i.e.} \qquad \mu \sim \frac{\lambda^{1/3}\, \nq^{1/3}}{\nc^{1/3} \, \nf^{1/3}} \ .
\ee

\begin{figure}[h!!!]
\begin{center}
 \includegraphics[width=0.8\textwidth]{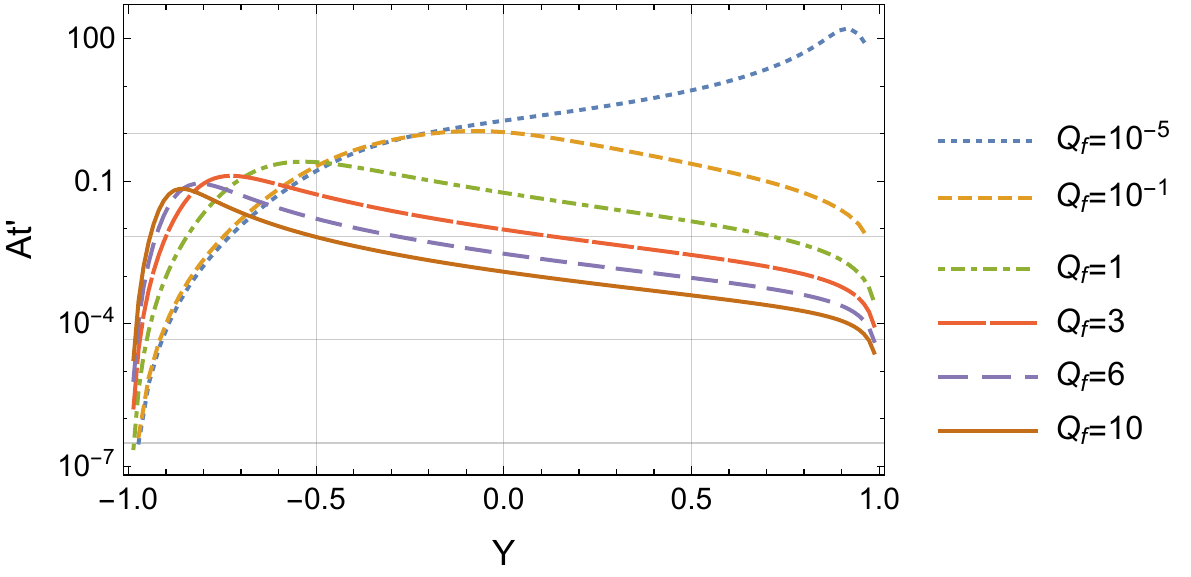}
\caption{ \label{fig.Atprimes}\small
Integrand of Eqn.~\eqref{eq.mubar} as a function of the compact radial coordinate for different values of $\rho$. Notice the logarithmic scale on the vertical axis. For large (small) values of $\rho$ the integrand   is mostly supported near the boundary (origin).}
\end{center}
\end{figure}
\begin{figure}[t]
\begin{center}
 \includegraphics[width=0.8\textwidth]{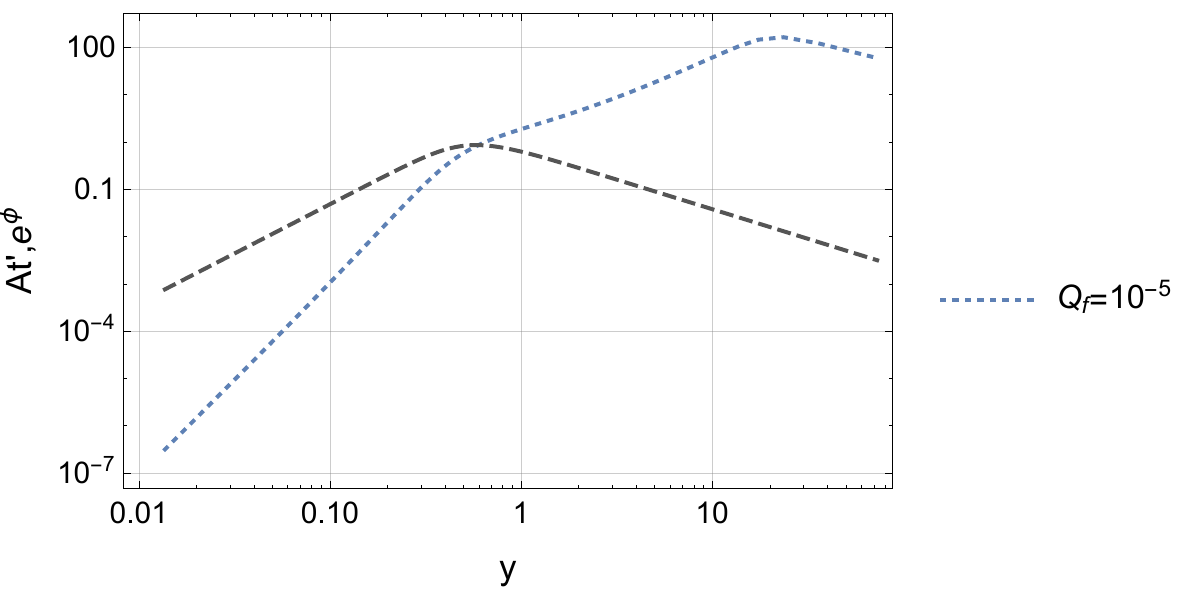}
\caption{ \label{fig.Atplargerho}\small
Integrand of Eqn.~\eqref{eq.mubar} (blue dotted curve) and radial profile of the dilaton (gray dashed curve) as a function of the radial coordinate $y$, for a large value of $\rho=1/12 \Qf^4$. The turning point of the dilaton takes place at a radial position dual to the energy scale $\ust$.}
\end{center}
\end{figure}
\begin{figure}[h!!!]
\begin{center}
 \includegraphics[width=0.8\textwidth]{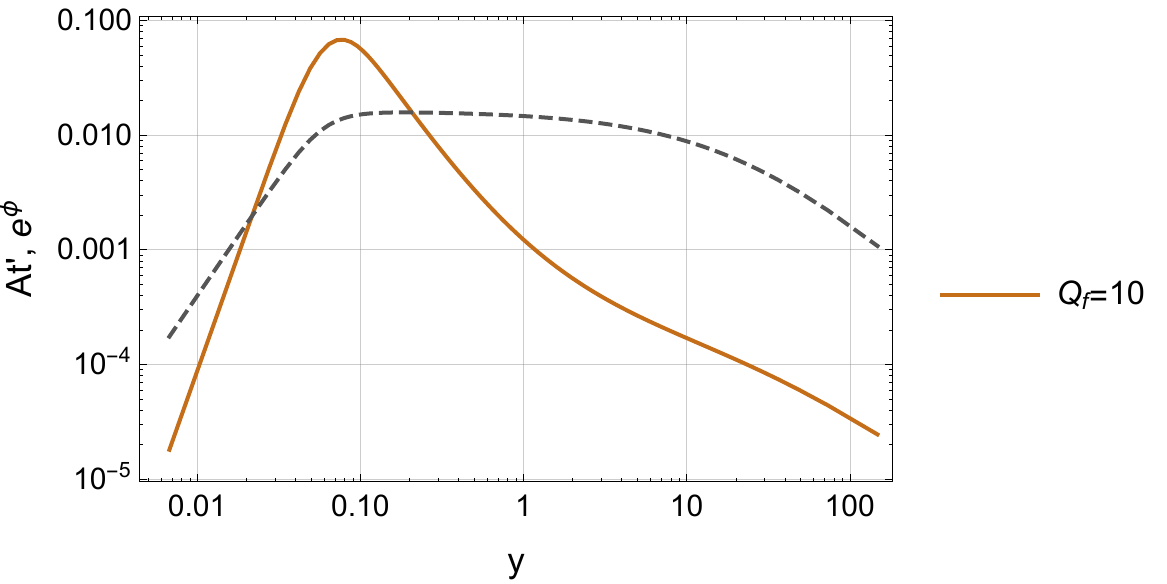}
\caption{ \label{fig.Atpsmallrho}\small
Integrand of Eqn.~\eqref{eq.mubar} (orange continuous curve) and radial profile of the dilaton (gray dashed curve) as a function of the radial coordinate $y$, for a small value of $\rho=1/12 \Qf^4$. The radial position at which $A_t'$ is peaked is dual to the energy scale $\uc$.}
\end{center}
\end{figure}

The exponents above can be understood  as follows. In 
Fig.~\ref{fig.Atprimes} we plot the integrand of Eqn.~\eqref{eq.mubar} as a function of the compact radial coordinate $Y$ defined in \eqref{eq.compacty}. We observe that the electric field $A_t'$ has support in different regions of the geometry  depending on the value of $\rho$. In particular, we see that for large or small values of $\rho$ it is mostly supported in the far UV or the deep IR, respectively (note the logarithmic scale on the vertical axis). To illustrate this point further, in Figs.~\ref{fig.Atplargerho} and \ref{fig.Atpsmallrho} we have superimposed the dilaton and the $A_t'$ profiles, as functions of the non-compact radial coordinate $y$, for our largest and  smallest available values of 
$\rho$ respectively. In the first case we see that the electric field is peaked at a scale much higher  than the scale $\ust$ at which the dilaton profile turns around (note the logarithmic scale on the horizontal axis). In the second case we see that $A_t'$ is peaked around the scale $\uc$ at which the geometry transitions between HVL and AdS$_4$, and has very little support in the region connecting AdS$_4$ and the D2-brane geometry. 
These observations suggest that we can estimate the chemical potential as follows.  

In the case of large $\rho$ we can approximate the integral in \eqref{eq.mubar} by the corresponding integral in the UV region in which the geometry is that of the D2-branes:
\be\label{eq.mubarlargerho}
\overline \mu \simeq \int_{\ls^2\, \ust}^\infty \sqrt{- \overline G_{tt} \overline G_{yy}}   
\frac{e^{\overline \phi} \,\overline G_{\Omega\Omega}^{-3/2}  
\rho }{\sqrt{ \overline G_{xx}^2 + 
e^{2 \overline \phi} \, \overline G_{\Omega\Omega}^{-3} 
\rho^2}}  \d \overline y + \cdots \ ,
\ee
where the dots stand for $\rho$-subleading corrections that include flavor effects near $y=\ls^2\,\ust$ where $A_t'$ is still  small. Performing the integral \eqref{eq.mubarlargerho} in the background \eqref{eqn.D2branes} we get
\be\label{eq.murhoonethird}
\overline \mu \simeq \frac{\Gamma\left(\frac{1}{6}\right) \Gamma\left(\frac{1}{3}\right) }{6 \, \sqrt{\pi}}\rho^{1/3} \simeq 1.4 \, \rho^{1/3}   \,.
\ee
Note that not just the 1/3 exponent but the numerical $1.4$ prefactor agree with the numerical fit in Fig.~\ref{fig:IRdof_chemicalPot}. This result coincides with the one obtained in the probe approximation  
\cite{Karch:2007br}, where one considers a single D6-brane in the background of a stack of D2-branes, with the probe brane wrapping an $\text{S}^3 \subset\text{S}^6$. From our discussion in Sec.~\ref{crossovers} we know that  the probe approximation fails to describe correctly the IR physics of the theory, but it is reliable in the UV. This presumably explains the agreement with our result \eqq{eq.murhoonethird}, since in the large-$\rho$ limit the dominant contribution to the chemical potential comes from the UV part of the geometry.

In the opposite limit, $\rho\to0$, the dominant contribution to the chemical potential comes from the region between the AdS$_4$ fixed point and the HVL geometry. Since $\nq$ is the only scale from the viewpoint of this fixed point, this means that the chemical potential is constrained by dimensional analysis to take the form 
\be
\mu \sim {\cal M}(\nf,\nc)\,\nq^{1/2} \qquad \mbox{i.e.} \qquad \overline \mu \sim \rho^{1/2} \,.
\label{indeed}
\ee
The prefactor ${\cal M}$  goes as ${\cal M}\sim \nf^{-1}$ from our result \eqref{eq.musmallrho}, which is a consequence of the scaling of the system discussed in Sec.~\ref{scalings}.
The contribution \eqq{indeed} is indeed dominant with respect to that of the region between the AdS$_4$ and the D2-brane geometries. To see this, we note that the asymptotic form \eqq{eqn.UVexp} of $\bfunc$, together with the rescalings of Sec.~\ref{scalings}, imply that $\bfunc \sim \rho$ at small $\rho$. It then follows from Eqn.~\eqref{eq.mubar}  that the  contribution to $\overline \mu$ from the region between the AdS$_4$ and the D2-brane geometries is linear in $\rho$.

With the results \eqq{eq.musmallrho} and \eqq{eq.mulargerho} in hand we now turn to the equation of state. We recall that the grand canonical potential $\Omega=\Omega(T,\mu)$ is equal to minus the pressure, $\Omega = -p$, and obeys
\be
\d \Omega = -s \,\d T - \nq \d \mu \,.
\ee
In contrast, the energy density is naturally a function $\epsilon(s,N_q)$ and obeys the first law of thermodynamics 
\be
\d \epsilon = T\, \d s + \mu \,\d \nq \,.
\ee
At zero temperature this relations imply that 
\begin{equation}
\frac{\d p}{\d \epsilon}=\frac{N_q}{\mu}\frac{\d\mu}{\d N_q}=
\frac{\rho}{\mu}\frac{\d\mu}{\d\rho} = 
\frac{\rho}{\overline \mu}\frac{\d \overline \mu}{\d\rho} 
\,.
\end{equation}
Consequently, for small values of $\rho$ we find $\epsilon=2p$, while for large values we get $\epsilon=3p + \mbox{const}$. We thus see that the speed of sound, $c_s^2 = \d p/\d \epsilon$, varies from the conformal value $c_s^2=1/2$ at small $\rho$ to $c_s^2=1/3$ at large $\rho$.

The fact that the chemical potential is a monotonically increasing function of the charge density, i.e.~that
\begin{equation}
\frac{\partial \overline \mu}{\partial \rho}>0\,,
\end{equation}
indicates that the system is locally thermodynamically stable against charge fluctuations.

\subsection{Entropy density at low temperature and IR degrees of freedom}
\label{entropysection}
We have seen that the IR geometry is always an HVL metric with fixed exponents $z, \theta$ regardless of the values of $\lambda, \nc, \nf$ and $\nq$ that define the theory. This symmetry implies that the entropy density at low temperatures must scale as in \eqq{prop}. However, the proportionality coefficient in this equation, which measures the number of low-energy degrees of freedom in the theory, does depend on $\lambda, \nc, \nf$ and $\nq$. Following Sec.~3.4 of \cite{Faedo:2014ana}, in this section we will determine this dependence analytically  for flows with very large or very small $\rho$, and numerically for the rest of them. In the equations below we will omit factors of $\ell_s$, since they can be simply reinstated by dimensional analysis, as well as purely numerical factors, since we are only interested in the parametric dependence of the entropy density on $\lambda, \nc, \nf$ and $\nq$.

It is straightforward to construct the non-zero-temperature generalization of the HVL solution \eqq{eqn.HVLifshitz}, which is a black hole with horizon radius $r_\mt{H}$ whose explicit form is given in Sec.~3.2 of \cite{Faedo:2014ana}. With this solution in hand one can compute the entropy density at low temperature from the area of the horizon. The  result is
\begin{align}
\label{entro}
&s \sim \frac{\nc^2}{\lambda^{3/2}} U_\mt{T}^{7/2}\ ,
\end{align}
with $U_\mt{T}$  the scale associated to the radius of the horizon via the relation \eqref{eqn.HVLifD2radialchange}.
The temperature is obtained from the Einstein-frame metric, $G^\mt{E}$, as
\be
\label{tcorrect}
T \sim \left. \frac{\partial_r (- G^\mt{E}_{tt})}{\sqrt{-G^\mt{E}_{tt}\, G^\mt{E}_{rr}}}\right|_{r\to r_\mt{H}} \sim\frac{\sqrt{C_\mt{t}} \, \nc \, U_\mt{T}^{35/2}}{\lambda^{43/10} \nq^{1/2}} \ .
\ee
Using \eqq{tcorrect} to eliminate $U_\mt{T}$ and substituting in \eqq{entro} we find
\be
s \sim C_\mt{t}^{-1/10} \lambda^{-16/25} \nc^{2} 
\left( \frac{\nq}{\nc^2} \right)^{1/10} \, T^{1/5} \,.
\ee  

It is crucial to note that the temperature is computed with respect to a time coordinate $t$ that is continuous along the flow and is normalized in the UV as in Eqn.~\eqq{eqn.D2branes}. In general, this normalization differs by a  rescaling from our (arbitrary) normalization  of the coordinate that appears in the IR metric  \eqq{eqn.HVLifshitz}, which we will denote here as $t_\mt{IR}$. Thus we have
\be\label{matching}
t = \frac{1}{\sqrt{C_\mt{t}}} t_\mt{IR} \ .
\ee
This is the origin of the $\sqrt{C_\mt{t}}$ factor in \eqq{tcorrect}.

\begin{figure}
\begin{center}
\includegraphics[width=0.55\textwidth]{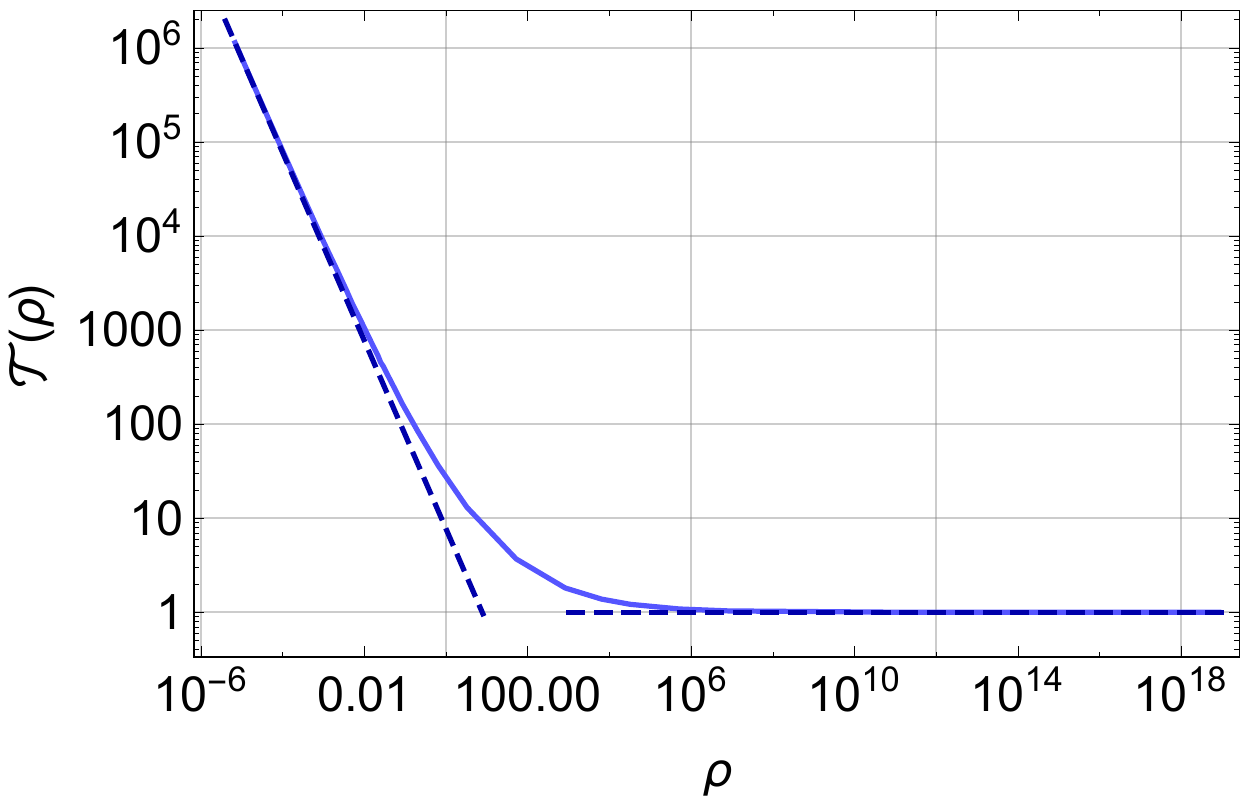}
\caption{\small \label{fig:IRdof}
The dimensionless function ${\cal T}(\rho)$ that enters the expression for the entropy density at low temperature, Eqn.~\eqq{all}. 
The dashed lines correspond to $1$ and a $1/\rho$ behavior.}
\end{center}
\end{figure}

To determine $C_\mt{t}$ we  approximate it by its value in the zero-temperature solutions presented in Sec.~\ref{numint}, which is accurate for low temperature. This means that we read $C_\mt{t}$ numerically by fitting the IR behavior shown in Eqn.~\eqq{ct}, where we have anticipated our approximation by naming the arbitrary constant with the same name as in \eqref{matching}. In particular, from our numerics we can determine the $\ratio$-dependence of the constant $C_\mt{t}$ up to an overall normalization. To determine this normalization we can study how the scalings in Sec.~\ref{scalings} affect Eqn.~\eqq{ct}. This exercise shows that, since the function $f_1$ is scale invariant, $C_\mt{t}$ must take the form
\be\label{eq.Ctscaling}
C_\mt{t} (\lambda, \nc, \nf, \nq) = \frac{\Qc^{4/15}}{\Qst^{16/3} \Qf^{20/3}} \overline{C}_\mt{t} (\rho) \ .
\ee
We have to complement the argument with a physical input that we will confirm below with an independent calculation: since in the limit $\rho\to\infty$ the effects of flavor are suppressed everywhere along the flow, in this limit the  entropy density must become $\nf$-independent. For the r.h.s.~of \eqref{eq.Ctscaling} to be $\nf$-independent we must have
\be\label{eq.Ctlargerho}
\lim_{\ratio\to\infty} \overline{C}_\mt{t}(\rho) \sim \rho^{-5/3} \qquad \mbox{i.e.} 
\qquad \lim_{\ratio\to\infty}  C_\mt{t}(\lambda,\nc,\nf,\nq) \sim  \frac{\nc^{14}}{ \nq^7 \, \lambda^{42/5}}  \ .
\ee
Therefore, we can write
\be
C_\mt{t} \sim  \frac{\nc^{14}}{ \nq^7\, \lambda^{42/5}} {\cal T}(\rho) \ ,
\ee
where we are omitting purely numerical factors and  $\cal T$ is a dimensionless function satisfying the condition $\lim_{\rho\to\infty}{\cal T} = 1$. We present this function, obtained from the numerical solutions at zero temperature, in 
Fig.~\ref{fig:IRdof}.
With this result in hand we can write the entropy density in terms of the $\ratio$-dependent quantity $\cal T$ as
\be
\label{all}
s\sim \nc^{2/5} \,\nq^{4/5}\,  \lambda^{1/5}\, T^{1/5}\, {\cal T}(\rho)\,.
\ee

The behavior in Eqn.~\eqref{eq.Ctlargerho} can be determined independently by requiring  that the norm of $\partial_t$ be continuous across the different transition scales, i.e.~by imposing that the time components of the UV and the IR metrics be related through \eqq{matching} at the transition scales. For flows with very large $\rho$ that transition directly from the D2 geometry to the HVL geometry we must demand continuity at  $\ust$. This exercise was done in \cite{Faedo:2014ana} with the result shown in  \eqref{eq.Ctlargerho}.

Using this strategy we can understand analytically also the $1/\rho$ behavior at small values of $\rho$ shown in Fig.~\ref{fig:IRdof}. In this limit the flow goes through an intermediate AdS region, and in principle one must demand  continuity of $C_\mt{t}$ across the two transition scales $\uf$ and $\uc$. However, the normalization of the time coordinate is parametrically the same in the D2 and in the AdS solutions, so it suffices to require continuity across $\uc$. This  results in 
\begin{equation}
\lim_{\rho\to0}C_\mt{t} \sim  \frac{\nc^{14}}{ \nq^7\, \lambda^{42/5}}  \frac{1}{\rho} \qquad \mbox{i.e.} \qquad \lim_{\rho\to0} {\cal T} \sim \frac{1}{\rho} \ ,
\end{equation}
and thus
\begin{align}
\lim_{\rho\to0}s \sim \frac{\nc^{3/5} \, \nq^{9/10}}{\nf^{2/5}}\, T^{1/5}\,.
\end{align}

From this behavior we observe that the number of IR degrees of freedom decreases with increasing $\nf$ as $s \sim \nf^{-2/5}$. One possible explanation for this behavior is that, as $\nf$ grows larger, the flow remains close to the AdS fixed point for a longer range of length scales, meaning that more degrees of freedom are integrated out before the flow finally reaches the HVL regime.

\section{Massive quarks}
\label{massive}
The qualitative behavior of the theory in the presence of a non-zero quark mass, $\Mq$, can be understood rather straightforwardly. This new scale in the gauge theory translates into a new radial position on the gravity side given by \cite{Faedo:2015ula}
\be\label{eqn.massdefinition}
\um =  2\pi\ls^2 \, \Mq \,.
\ee
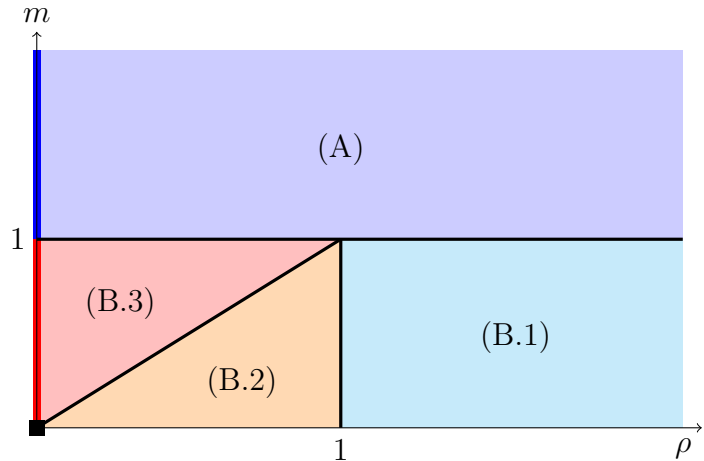
\begin{figure}\begin{center}
\begin{tikzpicture}[auto]
 
 \fill[cyan!20] (4,0)  -- (8.5,0) -- (8.5,2.5) -- (4,2.5);
 \draw (6.3,1.2) node {\large (B.1)};

 \fill[blue!20] (0,2.5) -- (8.5,2.5)  -- (8.5,5) -- (0,5) ;
 \draw (4,3.7) node {\large (A)};
 \fill[blue] (-0.05,5) -- (0.05,5) -- (0.05,2.5) -- (-0.05,2.5);
 
 \fill[orange!30] (0,0)  -- (4,0) -- (4,2.5);
 \draw (2.7,0.6) node {\large (B.2)};
 
 \fill[red!25] (0,0)  -- (4,2.5) -- (0,2.5);
 \draw (1.1,1.66) node {\large (B.3)};
 \fill[red] (-0.05,0) -- (0.05,0) -- (0.05,2.5) -- (-0.05,2.5);

 \fill (-0.1,-0.1) -- (0.1,-0.1) -- (0.1,0.1) -- (-0.1,0.1);

 \draw [very thick] (4,0) node [below] {\large $1$} -- (4,2.5);
 \draw [very thick] (0,2.5) node [left] {\large $1$} -- (8.5,2.5);
 \draw [very thick] (0,0)  -- (4,2.5);

 \draw [->] (0,0) -- (8.75,0) node [below left]  {\large $ \ratio$};
 \draw [->] (0,0) -- (0,5.25) node [above] {\large $\massratio$};
 
 \end{tikzpicture}
\caption{\small Qualitatively different classes of solutions in the presence of charge and massive quarks as a function of the dimensionless parameters $\rho$ and $m$ measuring the charge and the mass, respectively. The black square at the bottom left-hand corner represents the RG flow between the SYM and the CSM theories in the theory with massless quarks and no charge.} 
\label{fig.massiveRGflows}
\end{center}\end{figure}
The dynamics now depends non-trivially not only on $\rho$ but also on the dimensionless ratio between $\Mq$ and the flavor scale \eqq{uflavor}: 
\be\label{eqn.massratio}
\massratio \equiv  \frac{\Mq}{\uf} \,.
\ee

In the absence of charge the full solution for all values of $m$ was found in \cite{Faedo:2015ula}. The qualitative physics  can be simply understood by noticing that at energies below the quark mass the quarks decouple from the dynamics. On the gravity side this means that at radial positions below $\um$ the geometry is again given by the unflavored D2-brane geometry (albeit with renormalized parameters). Thus, if $m>1$, the quarks decouple from the theory before they can have a significant effect and the geometry is essentially given by the D2-brane geometry everywhere. In contrast, if $m<1$, then the theory exhibits quasi-conformal dynamics at energies $\Mq < U < \uf$. On the gravity side, the solution is given by the D2-brane geometry everywhere except for an approximate AdS$_4$ region between the radial positions $\um < u < u_\mt{flavor}$.

Consider now the theory in the presence of charge and massive quarks. 
Depending on the values of $\rho$ and $m$ there are four qualitative classes of flows,  represented in 
Fig.~\ref{fig.massiveRGflows}, corresponding to the following situations:
\begin{enumerate}[(A)]

\item If $\massratio > 1$ then the quarks decouple before they can have a significant effect on the dynamics, so the latter is qualitatively that of the unflavored theory with charge. The geometry is therefore well approximated by that of Ref.~\cite{Faedo:2014ana}, which corresponds to infinitely massive quarks. 

\item For $\massratio < 1$ we have to differentiate three cases, depending on the ordering of $\{m, \rho, 1\}$.

\begin{enumerate}[({B.}1)]

\item If $\massratio < 1 < \ratio$, Eqs.~\eqq{rhoratio} and \eqq{eqn.massratio}  imply that $\Mq < \ust$. Since $\rho > 1$ we are in the regime in which the charge density dominates the geometry, and since   $\Mq < \ust$ the quarks decouple at a scale below the relevant scale $\ust$. In other words, the quarks behave as effectively massless and the geometry is well approximated by that associated to the vertical line on the left-hand side of Fig.~\ref{fig.RGflows}.

\item For $\massratio< \ratio <1$, Eqs.~\eqq{rhoratio}, \eqq{hi} and \eqq{eqn.massratio} imply that $\Mq < \uc$. Since $\rho < 1$ we are in the regime in which the flow is close to the right-hand side of the triangle of Fig.~\ref{fig.RGflows}, and since   $\Mq < \uc$ the quarks decouple at a scale below the lowest relevant scale $\uc$. In other words, the quarks behave as effectively massless and the geometry is well approximated by one of the flows with an intermediate AdS$_4$ region on the right-hand side of Fig.~\ref{fig.RGflows}.

\item Finally, consider the case in which  $\ratio < \massratio < 1$ in such a way that $\uc < \Mq < \uf$. The flow starts off in the UV with the D2-brane geometry, exhibits an AdS$_4$ region at energy scales $\Mq < U < \uf$, becomes again well approximated by the D2-brane geometry at energy scales $\uc' < U < \Mq$, and finally enters a charge-dominated HVL region at energies $U < \uc'$. The coupling constant in the second D2 region is renormalized with respect to the UV coupling due to the integrating out of the quarks and (omitting some purely numerical factors) is given by \cite{Faedo:2015ula}
\be
\lambda ' \sim \left( 1 + \frac{1}{m} \right) \lambda \,.
\ee
Correspondingly, the scale $\uc'$ is given by \eqq{ucharge} with the replacement $\lambda \to \lambda'$, i.e.
\be
\uc' \sim \sqrt{\lambda'} \left( \frac{\nq}{\nc^2} \right)^{1/4} \sim
\Mq^{1/2}  \left( \frac{\nq}{\nf^2} \right)^{1/4} \,,
\ee
where the last expression is valid if $m$ is sufficiently smaller than 1. 
\end{enumerate}
\end{enumerate}

\section*{Acknowledgements}
We thank Aristomenis Donos, Pau Figueras, Matti J\"arvinen, Prem Kumar, Wilke van der Schee, Miquel Triana and Miguel Zilh\~ao for discussions. We are grateful to the Galileo Galilei Institute in Florence (AF, CP, JT), to the Centro de Ciencias de Benasque Pedro Pascual (AF, DM, CP, JT), to the Physics Departament at the University of California, Santa Barbara (DM), and to the Institute for Nuclear Theory, University of Washington (DM), for hospitality during part of this work. We are supported by grants 2014-SGR-1474, MEC FPA2013-46570-C2-1-P, MEC FPA2013- 46570-C2-2-P, CPAN CSD2007-00042 Consolider-Ingenio 2010, ERC Starting Grant HoloLHC-306605 and Maria de Maeztu Unit of Research Excellence distinction.
JT is also supported by the Juan de la Cierva program of the Spanish Ministry of Economy, by the Advanced ARC project ``Holography, Gauge Theories and Quantum Gravity'' and by the Belgian Fonds National de la Recherche Scientifique FNRS (convention IISN 4.4503.15). AK is supported by the Department of Atomic Energy, Govt. of India. He is also grateful for the warm hospitality at the University of Barcelona during the final stages of this work.

\appendix 
\section{Supergravity with sources}
\label{app_sources}
In this Appendix we will derive the equations of motion that we have solved in the main part of the paper. Our starting point is the type IIA supergravity action. This is the sum of a Neveu-Schwarz (NS) and a Ramond-Ramond (RR) sector,
\be
S_\mt{IIA} = S_\mt{NS} + S_\mt{RR} \,, 
\ee
respectively given by
\bse\label{eqn.IIAaction}
\bal
S_\mt{NS} & = \frac{1}{2\kappa^2} \int e^{-2\phi} \left( R*1 + 4\, \d \phi \wedge * \d \phi - \frac{1}{2} H \wedge * H \right) \ , \\
S_\mt{RR} & = \frac{1}{2\kappa^2} \, \frac{1}{2} \int \left( -  G_2 \wedge * G_2 - G_4 \wedge * G_4 +  H \wedge C_3 \wedge \d C_3 \right) \ ,
\end{align}
\ese
where the field strengths are defined as
\be
G_2 = \d C_\mt{1} \ , \qquad G_4 = \d C_3 - H \wedge C_\mt{1} \ , \qquad H = \d B \ .
\ee
The equations of motion that follow from this action for the NS and RR forms are
\bse\label{eqn.usualIIA}
\bal
\d * G_2 & = - H \wedge *G_4 \ , \\[2mm]
 \d * G_4 &  = - H \wedge G_4 \ , \label{use} \\[1mm]
\d \left(e^{-2\phi}*H\right) & = G_2 \wedge *G_4 + \frac{1}{2} G_4\wedge G_4 \ .
\end{align}
\ese
Note that in order to write the topological $G_4\wedge G_4$ term in the last equation in this form one must use \eqq{use}.

We will now consider the addition of electric and magnetic sources for $C_\mt{1}$ and $C_3$. For this purpose it is convenient to use the so-called `democratic formulation' \cite{Bergshoeff:2001pv} in which the Hodge duals of these potentials are treated as independent fields, thus making all the sources purely electric and hence mutually independent. The democratic formulation provides a pseudo-action instead of an action in the sense that the duality relations between Hodge potentials must be imposed by hand after obtaining the equations of motion. In this formulation the RR part of the action is replaced by
\be
S_\mt{RR}^\mt{democratic}  = \frac{1}{2\kappa^2} \, \frac{1}{2} \int \left( - \frac{1}{2} G_2 \wedge * G_2 - \frac{1}{2} G_4 \wedge * G_4 - \frac{1}{2} G_6 \wedge * G_6 - \frac{1}{2} G_8 \wedge * G_8 \right) \ ,
\ee
where the extra factors of $1/2$ account for the doubling of the degrees of freedom. The field strengths are defined as 
\be
G_n = \d C_{n-1} - H \wedge C_{n-3}   
\label{defG}
\ee
with $n=2,4,6,8$ and $C_{-1}=0$. The equations of motion that follow from the new pseudo-action for the RR fields are 
\bse
\label{eqn.democraticIIG}
\bal
\d * G_8 & = 0 \ , \\[1mm]
\d * G_6 & = - H \wedge * G_8 \ , \\[1mm]
\d * G_4 & = - H \wedge * G_6  \ , \\[1mm]
\d * G_2 & = - H \wedge * G_4  \ ,
\end{align}
\ese
whereas the equation for the $H$-field is
\bal
\d \left(e^{-2\phi}*H\right) & =  \frac{1}{2} \d \Big( C_\mt{1} \wedge  * G_4 +  C_3 \wedge  * G_6 +  C_5 \wedge  * G_8 \Big) \label{eqn.democraticIIAH}\nonumber \\[2mm]
& = \frac{1}{2} G_2 \wedge *G_4 + \frac{1}{2} G_4\ \wedge *G_6 + \frac{1}{2} G_6\ \wedge *G_8 \,. 
\end{align}
Note that in order to obtain this last equation one must make use of \eqref{eqn.democraticIIG}. The duality equations that must be imposed after the equations of motion have been obtained are 
\be\label{eqn.HodgeDemocratic}
G_8 = * G_2 \, \qquad G_6 = -*G_4 \ ,
\ee
where $**=-1$ for even forms. By means of these duality relations one may turn the equations of motion \eqq{eqn.democraticIIG} into the  set of Bianchi identities 
\be
\d G_n = H \wedge G_{n-2} \,,
\label{bibi}
\ee
which are explicitly solved by the definitions \eqq{defG} in terms of gauge potentials. 

We now consider the addition of smeared D6-branes with an arbitrary worldvolume gauge field turned on. The action for the D6-branes is the sum of a Dirac-Born-Infeld (DBI) term and a Wess-Zumino (WZ) term: 
\be
S_\mt{D6} = S_\mt{DBI}  + S_\mt{WZ}  \,.
\ee
In the following we will only need the explicit form of the WZ term:
\be\label{eqn.WZterm}
S_\mt{WZ} = \frac{1}{2\kappa^2} \int \frac{1}{2} \left( C_7 - C_5 \wedge \cF + \frac{1}{2} C_3 \wedge \cF \wedge \cF - \frac{1}{6} C_\mt{1} \wedge \cF \wedge \cF \wedge \cF  \right) \wedge \Gamma \ .
\ee
As above, the overall factor of $1/2$ is due to the doubling of degrees of freedom. We have absorbed the brane tension in the definition of $\Gamma$, which is a closed three-form that characterizes the distribution of D6-branes in the codimension-three space transverse to them. For example, for a single brane located at a particular point, specified by three coordinates $\zeta_\mt{D6}$, one would have 
\be
{\Gamma \sim \delta^{3}(\zeta-\zeta_\mt{D6})\, \d \zeta^1 \wedge \d \zeta^2 \wedge \d \zeta^3} \,. 
\ee
The field strength appearing in the WZ action is
\be
\cF = B + 2\pi\ell_s^2\, \d A \ ,
\ee
where $A$ is the Born-Infeld field living on the worldvolume of the D6-branes, and it satisfies the Bianchi identity 
\be
\d \cF = H \,.
\ee

The WZ term acts as a source for both the RR fields and the $H$-field, and hence it modifies their equations of motion. For the RR fields, we will see that, through the duality relations, this also leads to a modification of their Bianchi identities, and therefore to a modification of their very definition in terms of gauge potentials. For this reason we write the modified equations of motion in terms of new RR field strengths that we denote $F_n$. The equations for these RR fields  read
\bse\label{eqn.IIAdemocraticsourcedmod}
\bal
\d * F_8 & =  \Gamma \ , \\[1.9mm]
\d * F_6 & = - H \wedge * F_8 - \cF \wedge \Gamma \ , \label{eqn.G6eomsourced}\\[1mm]
\d * F_4 & = - H \wedge * F_6 + \frac{1}{2} \cF \wedge \cF \wedge \Gamma \ , \\[1mm]
\d * F_2 & = - H \wedge * F_4 - \frac{1}{6} \cF \wedge \cF \wedge \cF \wedge \Gamma \ .
\end{align}
\ese
Apart from the change of notation $G_n \to F_n$, these are just equations \eqq{eqn.democraticIIG} augmented with the extra source terms coming from the variation of \eqq{eqn.WZterm}.
We now see that in order to impose duality relations analogous to \eqq{eqn.HodgeDemocratic}, namely 
\be\label{eqn.Hodge_F}
F_8 = * F_2 \, \qquad F_6 = -*F_4 \ ,
\ee
we must modify the definitions of the RR field strengths with respect to those in \eqq{defG}. For example, upon use of \eqq{eqn.Hodge_F}, the first equation in \eqq{eqn.IIAdemocraticsourcedmod} reads
\be
\d F_2 =  - \Gamma \,,
\ee
which is obviously incompatible with $F_2=\d C_\mt{1}$. This problem is familiar from electromagnetism when both electric and magnetic sources are simultaneously present. In such a situation, in order to be able to define a gauge potential, one must modify the definition of the electric and magnetic fields in terms of this potential. In our case, the set of Bianchi identities that is obtained by using \eqq{eqn.Hodge_F} in \eqq{eqn.IIAdemocraticsourcedmod} is solved by the definitions
\bse\label{eqn.newforms}
\bal
F_2 & = G_2 +\gamma \ , \\[2mm]
F_4  & = G_4 + \cF \wedge \gamma \ , \\[1mm]
F_6 & = G_6 + \frac{1}{2} \cF \wedge \cF \wedge \gamma \ , \\[1mm]
F_8 & = G_8 + \frac{1}{6} \cF \wedge \cF \wedge \cF \wedge \gamma \,,
\end{align}
\ese
where $G_n$ are  defined  as in \eqq{defG} and $\gamma$ is a two-form satisfying 
\be\label{eqn.D6sourcesBI}
\d \gamma = - \Gamma \ .
\ee
Note that $\gamma$ always exists locally because $\Gamma$ is closed. 

The equation of motion for $H$ is \eqq{eqn.democraticIIAH} augmented by contributions from both the  D6-brane action and the definitions \eqref{eqn.newforms}. Upon the replacement $G_n \to F_n$ in the action, and taking into account the DBI and WZ terms this equation is now 
\bal
\d \left(e^{-2\phi}*H\right) & =  \frac{1}{2} \d \left( C_\mt{1} \wedge  * F_4 +  C_3 \wedge  * F_6 +  C_5 \wedge  * F_8 \right) \\
& \quad + \frac{1}{2} \left( * F_4 + \cF \wedge * F_6 + \frac{1}{2} \cF \wedge \cF \wedge *F_8 \right) \wedge \gamma \nonumber \\
& \quad + \frac{1}{2} \left( C_5 - \cF \wedge C_3 + \frac{1}{2} \cF \wedge \cF \wedge C_\mt{1} \right) \wedge \Gamma \nonumber \\
& \quad + \text{ DBI-terms} . \nonumber
\end{align}
Using  \eqq{eqn.newforms} and the equations of motion \eqq{eqn.IIAdemocraticsourcedmod}, the equation for the $H$-field can be simply written as
\begin{eqnarray}
\label{eqn.NSsourced}
\d \left(e^{-2\phi}*H\right)  &=&  \frac{1}{2} F_2 \wedge *F_4 + \frac{1}{2} F_4 \wedge *F_6 + \frac{1}{2} F_6 \wedge *F_8 \nonumber
\\[2mm] && + \, \text{DBI-terms}\ .
\end{eqnarray}

It is possible to write down the RR part of the supergravity action and the WZ term of the D6-brane action directly in terms of the modified field strengths $F_n$. In the democratic formulation these are given by 
\begin{eqnarray}
S_\mt{RR}^\mt{democratic}  &=& - \frac{1}{2\kappa^2} \, \frac{1}{2} \int \left( \frac{1}{2} F_2 \wedge * F_2 + \frac{1}{2} F_4 \wedge * F_4 + \frac{1}{2} F_6 \wedge * F_6 - \frac{1}{2} F_8 \wedge * F_8 \right) \,,
\\[2mm]
S_\mt{WZ} &=&  - \frac{1}{2\kappa^2} \, \frac{1}{2} \int  
\left( F_8 - F_6\wedge \cF + \frac{1}{2} F_4 \wedge \cF \wedge \cF -\frac{1}{6} F_2 \wedge \cF \wedge \cF \wedge \cF \right) \wedge \gamma \,.
\,\,\,\,\,\,\,\,\,\,\,\,\,\,\,\,\, 
\end{eqnarray}

We finally return to the non-democratic formulation and note that, in the presence of sources, the RR part of the supergravity action and the WZ term of the D6-brane action can be written directly in terms of the modified field strengths $F_2$ and $F_4$ alone: 
\begin{eqnarray}
S_\mt{RR}  &=& - \frac{1}{2\kappa^2} \, \frac{1}{2} \int \left( 
 F_2 \wedge * F_2 +  F_4 \wedge * F_4 - \frac{1}{2} H \wedge C_3 \wedge \d C_3 
\right) \,,
\\[2mm]
S_\mt{WZ} &=&  - \frac{1}{2\kappa^2} \, \int  
\left(  \frac{1}{2} F_4 \wedge \cF \wedge \cF -\frac{1}{6} F_2 \wedge \cF \wedge \cF \wedge \cF  - \frac{1}{6} \cF \wedge \cF \wedge \cF \wedge \gamma \right) \wedge \gamma \,.
\,\,\,\,\,\,\,\,\,\,\,\,\,\,\,\,\, 
\end{eqnarray}
By  varying the resulting complete action with respect to the dilaton, the metric and the RR and NS potentials, one obtains complete set of equations that we have solved in this paper and which we collect here. The equations of motion for the RR and NS forms are
\bse
\bal
\d * F_4 & = - H \wedge  F_4 + \frac{1}{2} \cF \wedge \cF \wedge \Gamma \ , \\
\d * F_2 & = - H \wedge * F_4 - \frac{1}{6} \cF \wedge \cF \wedge \cF \wedge \Gamma \ , \\
\d \left(e^{-2\phi}*H\right)  & =  F_2 \wedge *F_4 + \frac{1}{2} F_4 \wedge F_4 + \text{DBI-term} \ .
\end{align}
\ese
Since the gauge field $A$ living in the worldvolume of the D6-branes can be seen as a St\"uckelberg vector field for the NS two-form $B$, the equation of motion for $A$ follows from the equation of motion for $B$ and it takes the form
\be
\d \left[  F_2 \wedge *F_4 + \frac{1}{2} F_4 \wedge F_4+ \text{DBI-terms} \right] = 0 \ .
\ee
The equation of motion for the dilaton is given by
\be
R*1+4\d*\d \phi - 6 \d\phi \wedge * \d \phi - \frac{1}{2} H \wedge * H + \text{DBI-terms} = 0
\ee
Finally, the Einstein equations read
\bal
R_{\mu\nu} & = -   2 \nabla_\mu \nabla_\nu \phi + \frac{1}{4} H_{\mu\rho\sigma}{H_\nu}^{\rho\sigma} \\[1mm]
& \qquad  + e^{2\phi} \left[ \frac{1}{2}\left( F_2 \right)_{\mu\rho} \left( F_2 \right)_\nu{^\rho}+ \frac{1}{12}\left( F_4 \right)_{\mu\rho\sigma\alpha} \left( F_4 \right)_\nu{^{\rho\sigma\alpha}} - \frac{1}{4}G_{\mu\nu} \left( F_2 \lrcorner F_2 + F_4 \lrcorner F_4 \right)  \right] 
\nonumber \\[2mm]
& \qquad + \text{DBI-terms}  \,, \nonumber
\end{align}
where for any two $n$-forms $\omega, \xi$ we have defined 
\be
\omega \lrcorner \xi = \frac{1}{n!} \omega_{M_1 \ldots M_n}
\xi^{M_1 \ldots M_n} \,.
\ee
In the bulk of the paper we have written
\be
\Gamma = 2\kappa^2\, T_\mt{D6} \, \Xi \ , \qquad \qquad 
\gamma = \Qf\, J \ .
\ee

\section{Equations of motion}
\label{eomG}
In this section we will write down explicitly the full set of equations of motion for the overlined variables introduced in Sec.~\ref{scalings}. Eq.~\eqq{eqn.Atequation} for the time component of the $U(1)$ vector field takes the form  
\eqq{eqn.AtequationBis} in terms of these variables.
This could be substituted in  the remaining equations to eliminate the dependence on $\At'$. However, we will follow the alternative route of eliminating  $\At'$ at the level of the action by performing a Legendre transform that modifies the contribution of the sources in \eqref{eqn.totalaction} to
\be\label{eqn.legendretransform}
S_\mt{sources}- \int \frac{\partial S_\mt{sources}}{\partial \At'}\At' = - \frac{\Qf^5}{2\kappa^2} \int e^{-\phi} \,G_{\Omega\Omega}^{-3/2} \,\sqrt{1+\left( \frac{e^\phi (\ratio+\bfunc)}{G_{xx}\,G_{\Omega\Omega}^{3/2}} \right)^2} *1 + T_{D6} \int C_7 \wedge \Xi\ .
\ee
To arrive at \eqref{eqn.legendretransform} we have explicitly used the fact that the NSNS form vanishes, $B=0$. Otherwise there would be an extra term proportional to $B_{yt}$, coming from the coupling in the WZ part of the D6-action between $B$ and and the worldvolume gauge field, $A$. In our case it is consistent to set $B=0$ in the action at this stage because the sourcing due to the strings has already been taken into consideration explicitly.

After elimination of $A_t'$, the equations of motion take the following form. The equation for the function $\bfunc$ appearing in the RR fluxes is 
\be
\partial_y \left( \sqrt{\frac{-G_{tt}}{G_{yy}}} \, \frac{\bfunc'}{G_{xx}} \right) - 12 \frac{\sqrt{-G_{tt}\, G_{yy}}}{G_{xx}\,G_{\Omega\Omega}}\, \bfunc -  \frac{3\, e^\phi  \sqrt{-G_{tt}\,G_{yy}} \,(\ratio+\bfunc)}{\sqrt{G_{xx}^2\,G_{\Omega\Omega}^3 + e^{2\phi} 
\, (\ratio+\bfunc)^2}}= 0 \ .
\ee
The dilaton equation of motion becomes
\bal
0 =& \phi''  + \log'\left[ e^{-2\phi} \frac{\sqrt{-G_{tt}}\,G_{xx}\,G_{\Omega\Omega}^3}{\sqrt{G_{yy}}} \right] \phi' +  \frac{ e^{2\phi}\, \bfunc'^2}{G_{xx}^2\, G_{\Omega\Omega}^3}  +\frac{1}{2} {\cal F}_{1,-9,-3,3}(y) \\[2mm]
&   - 3\, e^\phi \frac{ G_{yy}\, \left( 3\, G_{xx}^2 \, G_{\Omega\Omega}^3 +2\, e^{2\phi} \, (\ratio+\bfunc)^2 \right)}{G_{xx}\, G_{\Omega\Omega}^3 \sqrt{G_{xx}^2 \, G_{\Omega\Omega}^3 + e^{2\phi} \, (\ratio+\bfunc)^2}  }  \nonumber \,,
\end{align}
where for notational convenience we have defined
\be
{\cal F}_{A,B,C,D}(y)= e^{2\phi} \frac{ G_{yy}}{2\,G_{xx}^2 G_{\Omega\Omega}^6} \Big[ G_{xx}^2 \left( A+ B\, G_{\Omega\Omega}^4 \right) + 16 \, G_{\Omega\Omega}^2 \left(9C\, \ratio^2 \,G_{\Omega\Omega}^4 + D\, \bfunc^2 \right) \Big] \ .
\ee
Finally, Einstein equations reduce to the following set of second order differential equations
\bse\bal
0 \, = \,\, & G_{tt}''  + \log'\left[ e^{-2\phi} \frac{G_{xx}\,G_{\Omega\Omega}^3}{\sqrt{-G_{tt}\, G_{yy}}} \right]G_{tt}' +  \left( \frac{2 \, e^{2\phi}\,\bfunc'^2}{G_{xx}^2\, G_{\Omega\Omega}^3}  - {\cal F}_{1,3,1,3}(y) \right) G_{tt}\\[2mm]
&   - 6\, e^\phi \frac{ G_{yy}\, G_{tt}\, \left( G_{xx}^2 \, G_{\Omega\Omega}^3 +2\, e^{2\phi} \, (\ratio+\bfunc)^2 \right)}{G_{xx}\, G_{\Omega\Omega}^3 \sqrt{G_{xx}^2 \, G_{\Omega\Omega}^3 + e^{2\phi} \, (\ratio+\bfunc)^2}  } \nonumber \ , \\[6mm]
0 \, = & \, \, G_{xx}'' + \log'\left[ e^{-2\phi} \frac{\sqrt{-G_{tt}}\,G_{\Omega\Omega}^3}{\sqrt{G_{yy}}} \right] G_{xx}' + \left( \frac{2 \, e^{2\phi}\,\bfunc'^2}{G_{xx}^2\, G_{\Omega\Omega}^3}  + {\cal F}_{-1,-3,1,3}(y) \right) G_{xx}\\[2mm]
&  -6\, e^\phi   \frac{ G_{yy}\, G_{xx}^2}{ \sqrt{G_{xx}^2 \, G_{\Omega\Omega}^3 + e^{2\phi} \, (\ratio+\bfunc)^2}  }  \nonumber \ , \\[6mm]
0 \, = & \, \, G_{\Omega\Omega}''  + \log'\left[ e^{-2\phi} \frac{\sqrt{-G_{tt}}\, G_{xx}\,G_{\Omega\Omega}^2}{\sqrt{ G_{yy}}} \right] G_{\Omega\Omega}'  + {\cal F}_{1,-1,-1,1}(y)\, G_{\Omega\Omega}  -10 \, G_{yy}  \ ,
\end{align}\ese
together with the first order constraint
\bal\label{eqn.constraint}
0= &\, \phi' \log' \left[ e^{-\phi}\sqrt{-G_{tt}}\, G_{xx}\, G_{\Omega\Omega}^3 \right] + \log' \left[ G_{\Omega\Omega}^{-3/2} \right] \log' \left[ \sqrt{-G_{tt}}\, G_{xx}\, G_{\Omega\Omega}^{5/4} \right] \\[2mm]
&  + \log' \left[ G_{xx}^{-1/2} \right] \log' \left[ \sqrt{-G_{tt}}\, G_{xx}^{1/4} \right] + \frac{e^{2\phi}\,\bfunc'^2}{2 \, G_{xx}^2\, G_{\Omega\Omega}^3} + \frac{15}{2}\frac{G_{yy}}{G_{\Omega\Omega}}-\frac{1}{4} {\cal F}_{1,3,1,3}(y) \nonumber \\[2mm]
&   - 3 \frac{e^\phi \,G_{yy}}{G_{xx}\, G_{\Omega\Omega}^3} \sqrt{G_{xx}^2\,G_{\Omega\Omega}^3 + e^{2\phi} \, (\ratio+\bfunc)^2}  \ . \nonumber
\end{align}
An effective one-dimensional  Lagrangian giving rise to these equations of motion can be derived from the constraint as
\be
\mathcal L_\mt{eff} = H- \sum_i \psi_i' \, \frac{\partial H}{\partial \psi_i'} \ ,
\ee
 with $\psi_i$ running over all fields and $H$ the r.h.s of \eqref{eqn.constraint}.

\section{Numerical construction of solutions}
\label{num}
With the ansatz \eqq{eq:ansatz} the equation of motion for $f_2$ can be solved algebraically with the result 
 \begin{equation}
 \label{f2fun}
 f_2=\frac{e^{-4 \chi}
  f_1\,g^2 (-1500 e^{4 \chi} y^8 +  50 \mathcal Z \,e^{6 \chi}y^4\,g + (e^{12 \chi} \Qc^2 \Qst^2 +  75 e^{8 \chi} \Qf^2 y^6 + 625 y^8 + 
      1200 e^{4 \chi} y^2 {\cal B}^2) g^2)}{25 y^4 (y^5 g f_1' (4 y g'-7g) +f_1 (4 g^4 {\cal B}'^2 + 38 y^5 g g' - 8 y^6 g'^2 +y^4 g^2 (-35 - 6 y \chi' + 12 y^2 \chi'^2))))}\,,
 \end{equation}
 where we have defined
 \be
 \mathcal Z^2 = 144 Q_\mt{f}^2 y^6 + (\Qc \Qst + 12 \Qf \mathcal B)^2 g^2 \,.
 \ee
 The gauge field $A_t$ on the D6-branes can be solved in terms of the other functions and is given by
 \be 
 \label{att}
 2 \pi \alpha' A_t'=\sqrt{\frac{f_1}{f_2}}\frac{e^\chi (\Qc \Qst + 12 \Qf {\cal B}) 
  g}{\mathcal Z }\,.
\ee
 The remaining functions obey the  following  set of  four, coupled, second-order differential equations     
     \begin{align}\label{four}
0=&{\cal B}''+\log'[\sqrt{f_1 f_2} e^{-\chi}]\,{\cal B}'-\frac{12 {\cal B}}{y^2 f_2} -\Qf\frac{3 \,e^{2\chi}\,g\, (\Qc \Qst + 12 \Qf {\cal B})}{f_2\, \mathcal Z }\,,\nonumber\\
0=&g''-\frac{g^3{\cal B}'^2}{2\, y^6}+\frac{1}{2}\log'\left[\frac{f_2 \,y^{19/2}}{g^4}\right]-\frac{7}{8}\frac{g}{y}\log'[y^5\,f_2 \,e^{-20\chi/7}]-\chi'\, \frac{(g\, e^{3\chi/2})'}{e^{3\chi/2}}+\frac{15 \,g}{2\, y^2\, f_2} -\frac{g^2\,e^{2\chi}\,\mathcal Z }{4\, y^6\, f_2}\nonumber\\
&-\frac{g^3\,e^{-4 \chi}}{200\, y^{10}\, f_2}(625\, y^8+1200\, y^2\,e^{4 \chi}{\cal B}^2+\Qc^2\Qst^2\,e^{12 \chi}+75 \,y^6 \Qf^2\,e^{8 \chi})\nonumber\\
0=& \chi''-\frac{1}{2}\log'\left[\frac{f_1\, f_2}{g^4}\right]\log'[y^{1/4}e^{-\chi}]-\chi'^2-\frac{5}{4 \,y}\log'[y\, e^{-5\chi}]-\frac{5}{y^2 f_2}\nonumber\\
&+\frac{g^2\,e^{-4 \chi}}{100\, y^{10}\, f_2}(625\, y^8+400\, y^2\,e^{4 \chi}{\cal B}^2-\Qc^2\Qst^2\,e^{12 \chi}-25 \,y^6 \Qf^2\,e^{8 \chi})\nonumber\\
0=&f_1''-f_1'\log'[y^{-29/4}\,f_1^{1/2} e^\chi]+\frac{(y^{5/2} f_1)'}{2\, y^{5/2}}\log'\left[\frac{f_2}{g^4}\right]-\frac{g\,f_1\,e^{2 \chi}}{y^6\,f_2\,\mathcal Z }\left({\mathcal Z}^2-72 Q_\mt{f}^2 y^6 \right)-\frac{5\, f_1}{2\, y}\chi'\nonumber\\
&-\frac{2\, g^2 f_1}{y^6}{\cal B'}^2+\frac{25 f_1}{2\, y^2}-\frac{g^2\,f_1\,e^{-4 \chi}}{50\, y^{10}\, f_2}(625\, y^8+1200\, y^2\,e^{4 \chi}{\cal B}^2+\Qc^2\Qst^2\,e^{12 \chi}+75 \,y^6 \Qf^2\,e^{8 \chi})
    \end{align}
   
The boundary conditions that must be imposed are as follows. 
 In the UV, as $y\to \infty$, we demand that we have the asymptotic behavior given by
\bse
\label{eqn.UVexp}
\bal
{\cal B} & = \frac{\Qc \, \Qst}{\Qf} \Bigg( 0 - \frac{1}{40} \frac{\Qf}{y} +  \frac{41}{400} \frac{\Qf^2}{y^2} + {\cal B}_3 \frac{\Qf^3}{y^3} + \cdots + \log\left[\frac{\Qf}{y}\right] \left(  \frac{143}{840}\frac{\Qf^3}{y^3} + \cdots \right) \Bigg) \ , \\
f_1 & = 1 + {\cal V}_1 \frac{\Qf^5}{y^5} + \cdots + \log\left[\frac{\Qf}{y}\right] \left(  \frac{-143 \Qc^2 \Qst^2}{42000}\frac{\Qf^2}{y^{10}} + \cdots \right) \ , \\
g & = 1 - \frac{13}{10}  \frac{\Qf}{y} 
+\cdots+  {\cal V}_2   \frac{\Qf^5}{y^5} + \cdots + \log\left[\frac{\Qf}{y}\right] \left(  \frac{-143 \Qc^2 \Qst^2}{120000}\frac{\Qf^2}{y^{10}} + \cdots \right)\ , \\
\chi & = 0 - \frac{17}{30}  \frac{\Qf}{y} 
+ \cdots +  {\cal V}_3   \frac{\Qf^{10}}{y^{10}} + \cdots+ \log\left[\frac{\Qf}{y}\right] \left(  \frac{5291 \Qc^2 \Qst^2}{2520000}\frac{\Qf^2}{y^{10}} + \cdots \right)\ , 
\end{align}\ese
The most important thing to notice is that this implies that the solution approaches the D2-brane solution \eqref{eqn.D2branes}. The expansion \eqref{eqn.UVexp} is specified by four dimensionless parameters ${\cal B}_3$, ${\cal V}_1$, ${\cal V}_2$ and ${\cal V}_3$ holographically dual to the expectation values of the operators dual to $\cal B$, the stress tensor, $\mbox{Tr} F^2$ and $\mbox{Tr} F^4$ respectively.

In the IR, as $y\to0$,  we demand that the fields approach their values in the  HVL solution 
\eqq{eqn.HVLifshitz}.  In the radial coordinate that we have chosen this implies the following expansion
\bse
\label{eqn.IRexp}
\begin{align}
\cal B &= C_\mt{B} \, y^{\Delta_3}- \frac{3 \cdot 5^{2/3}\Qf}{(14\, \Qc \Qst)^{1/3}} \,y^{10/3} +\cdots\, ,\\
f_1 & = C_\mt{t}  \, \frac{\Qst^{16/3}}{\Qc^{4/15}} \,y^{20/3} \left( 1+\frac{1}{11}(87+\sqrt{793})c_\mt{1} \, y^{\Delta_1 }  + \frac{1}{11}(87-\sqrt{793})c_\mt{2} \, y^{\Delta_2} +\cdots\right),  \label{ct} \\
 g & = \frac{5^{1/6} \, 14^{2/3} }{11^{1/2} (\Qc \Qst)^{1/3} } \, y^{4/3} \left( 1+\frac{1}{22}(21+\sqrt{793})c_\mt{1}\, y^{\Delta_1 }  + \frac{1}{22}(21-\sqrt{793})c_\mt{2}\, y^{\Delta_2} +\cdots\right)\ ,  \\
e^\chi & = \frac{10^{1/3}}{ (7 \Qc \Qst)^{1/6} } \, y^{2/3} \left( 1+c_\mt{1} \, y^{\Delta_1 }  + c_\mt{2} \, y^{\Delta_2} +\cdots\right)\,.
\end{align}\ese
where  we have reabsorbed a purely numeric factor in $C_\mt{t}$ with respect to the one in \eqq{eqn.HVLifshitz} and
\begin{eqnarray}
\Delta_1&=&\frac{5}{6}(-3 + \sqrt{ \frac{3}{11} ( 381 - 4 \sqrt{793} ) }) \\
\Delta_2&=&\frac{5}{6}(-3 + \sqrt{ \frac{3}{11} ( 381 + 4 \sqrt{793} ) } ) \\
\Delta_3&=&{\frac{5}{6} \left( -1 + \sqrt{ \frac{149}{5} } \right) } \,.
\end{eqnarray}
Note that the IR expansion is completely determined in terms of four constants $c_\mt{1}$, $c_\mt{2}$, $C_\mt{t}$ and $C_\mt{B}$.

\section{Spectrum of fluctuations around AdS$_4$}
\label{spectrum}
In this appendix we will compute the spectrum of fluctuations around the AdS$_4$ fixed point present in Fig.~\ref{fig.RGflows}. Using  the standard AdS/CFT dictionary this will allow us to find the dimension of the dual operators in the gauge theory. 

The internal geometries we considered admit a NK structure, which is a particular instance of $SU(3)$ structure in six dimensions. Reduction of massive type IIA supergravity on this type of manifolds was considered in \cite{KashaniPoor} and gives rise to an $\mathcal{N}=2$ supergravity in four dimensions. When Romans' mass is non-vanishing, this supergravity admits the $\mathcal{N}=1$ AdS solution first found in \cite{Behrndt}. If furthermore one allows for explicit (smeared) D6-brane sources then a new AdS vacuum exists: the IR fixed point in the flows of \cite{Faedo:2015ula} represented by one of the vertices in Fig.~\ref{fig.RGflows}. In the following we will compute the spectrum of fluctuations around this solution. 

In order to respect the NK structure, we have to expand the supergravity fields using the globally defined forms. For instance, the NS-form is expanded as
\begin{equation}
B\,=\,b_2+b_0\,J \,,
\ee
which implies
\be
H\,=\,\dd b_2+\dd b_0\wedge J+3\,b_0\,{\rm Im}\,\Omega\,,
\end{equation}
where $b_0$ and $b_2$ are respectively a scalar and a two-form living in the four-dimensional external spacetime. Similarly, the RR one and three-form potentials take the form
\begin{eqnarray}\label{RRexpansion}
C_\mt{1}&=&\tilde{a}_1\,,\nonumber\\[2mm]
C_3&=&a_3+a_1\wedge J+a_0\,{\rm Im}\,\Omega+\tilde{a}_0\,{\rm Re}\,\Omega\,,
\end{eqnarray}
where again $a_i$ and $\tilde{a}_i$ are $i$-forms in four dimensions. Notice that three-forms like $a_3$ are non-dynamical in four dimensions and should eventually be dualized to a constant. Up to this point the ansatz coincides with the one in \cite{KashaniPoor} (which in addition  considered  a non-vanishing Romans' mass). However, in our case the explicit sources modify the Bianchi identities and induce new couplings in the reduced action. We then choose to solve the Bianchi identities as\footnote{This choice is different from that used in Appendix \ref{app_sources}, since it is more convenient for the dimensional reduction. Of course, the two choices are related by a gauge transformation given by 
\be
C_3^\mt{there} = C_3^\mt{here} + \cF \wedge C_\mt{1} \ .
\ee}
\begin{eqnarray}
F_2&=&\dd C_\mt{1}+\Qf\,J\,,\nonumber\\[2mm]
F_4&=&\dd C_3+\cF\wedge F_2\,,
\end{eqnarray} 
where $\cF=2\pi\ell_s^2 \, \dd A+B$, with $A$ denoting the gauge field on the D6-branes. Since there is no globally-defined one-form in our construction, the gauge field on the branes will not have internal components. The final piece of the ansatz is the metric in string frame, which reads
\begin{equation}
\dd s_\mt{string}^2\,=\,e^{\frac{\phi}{2}}\left(e^{-6U}\,\dd s_4^2+e^{2U}\,\dd s^2_6\left({\rm NK}\right)\right)\,
\end{equation}
where the dilaton $\phi$ is assumed to depend only on the external coordinates $\dd s^2_4$. The internal metric $\dd s_6^2$ is that of a nearly K\"ahler manifold normalized to have curvature $R_{\text{\tiny{NK}}}=30$, and we have tuned the factors so that the reduced action is in Einstein frame. 

The properties of the globally defined forms, in particular closure of the set of forms under exterior derivative, wedging and Hodge duality, ensures that the ansatz is consistent and one can reduce the equations of motion to a four-dimensional system. We begin by analyzing the scalar-plus-gravity sector of this system, which can be  described by the action\footnote{The scalar $a_0$ in (\ref{RRexpansion})  is axionic in nature, its only coupling being a St\"uckelberg one to a vector. It will therefore be considered together with the vector fluctuations. Likewise, the two-form $b_2$ could be dualized to a scalar. This also leads to a St\"uckelberg coupling to another vector and we thus postpone the discussion.}
\begin{eqnarray}
S_{\text{\tiny scalar}}&=&\frac{1}{2\kappa_4^2}\,\int\,\left[R*1-24\,\dd U\wedge*\dd U-\frac12\dd\phi\wedge*\dd \phi-\frac32e^{-4U-\phi}\dd b_0\wedge*\dd b_0\right.\nonumber\\[2mm]
&&\left.\qquad\qquad-2\,e^{-6U+\frac{\phi}{2}}\dd \tilde{a}_0\wedge*\dd \tilde{a}_0-V*1\right]\,,
\end{eqnarray}
with a potential
\begin{eqnarray}
V&=&\frac12e^{-18U-\frac{\phi}{2}}\left(12\,b_0\,\tilde{a}_0+3\Qf b_0^2- \Qc \right)^2+6\,e^{-14U+\frac{\phi}{2}}\left(2\tilde{a}_0+ \Qf b_0\right)^2\nonumber\\[4mm]
&+&\frac32 \Qf^2 e^{-10U+\frac32\phi}-30\,e^{-8U}+18\,b_0^2e^{-12U-\phi}+12\, \Qf e^{-9U+\frac34\phi}\,.
\end{eqnarray}
The potential in \cite{KashaniPoor} for vanishing Romans' mass can be recovered by setting $\Qf=0$ to zero. Another known limit is to fix $b_0=\tilde{a}_0=0$ so that the potential in \cite{Faedo:2015ula} is reproduced with the appropriate change of variables. The constant parameter $\Qc$, proportional to the number of color branes, is roughly the Hodge dual of the three-form $a_3$.

This potential admits an AdS vacuum solution preserving $\mathcal{N}=1$ at the values of the scalars
\begin{equation}\label{AdSsolution}
\begin{array}{rclcrcl}
e^{4U+\phi}&=&\frac19\,\frac{\Qc}{\Qf}\,,&\qquad\qquad\qquad&\tilde{a}_0&=&0\,,\\[4mm]
e^{-6U+\frac{\phi}{2}}&=&\frac94\,\frac{1}{\Qc\, \Qf}\,,&\qquad\qquad\qquad&b_0&=&0\,,
\end{array}
\end{equation}
which is the IR fixed point in the RG flows of \cite{Faedo:2015ula} and one of the vertices in  Fig.~\ref{fig.RGflows}. The radius of this solution is 
\be
L=\frac{4}{9\sqrt{3}}\left(\Qc^3\Qf\right)^{\frac14} \,. 
\ee
It is now straightforward to compute the spectrum of fluctuations around this fixed point, together with the dimension of the dual operator in the CFT, with the result summarized in Table~\ref{scalarspectrum}.

\begin{table}[h]
\resizebox{0.95\textwidth}{!}{\begin{minipage}{\textwidth}
\begin{center}
\begin{tabular}{lcc}
\hline
\rowcolor{gray}	Mass Eigenstate&$ \qquad m^2L^2$&$\qquad\Delta$\\[6pt]
\hline
$36\,\delta U+\delta\phi$&$\qquad18$&$\qquad6$\\[6pt]
\rowcolor{gray}	$2\,\delta\tilde{a}_0-3\Qf\,\delta b_0$&$\qquad10$&$\qquad5$\\[6pt]
$4\,\delta U-3\,\delta\phi$&$\qquad\frac{22}{9}$&$\qquad\frac{11}{3}$\\[6pt]
\rowcolor{gray}	$\delta\tilde{a}_0+2\Qf\,\delta b_0$&$\quad-\frac89$&$\qquad\frac83$\\[6pt]
\hline
\end{tabular}
\end{center}
\caption{\small Spectrum of scalars around the supersymmetric AdS$_4$ solution.}
\label{scalarspectrum}
\end{minipage} }
\end{table}

In the flow from three-dimensional SYM to a CSM theory found in \cite{Faedo:2015ula}, the IR fixed point is approached along the irrelevant directions corresponding to the operators of dimensions $\Delta=6$ and $\Delta=11/3$. These scalar operators are also the only ones active in the charged flows presented in the bulk of the paper. As explained in Sec.~\ref{numint}, the fact that there are no relevant operators turned on at the AdS$_4$ fixed point is an important consistency condition. 

Finding the spectrum of vectors is more involved because several dualizations are needed. We start with the action for the vectors at quadratic order, which is all we need for the study of the spectrum:
\begin{eqnarray}
S_{\text{\tiny{vector}}}&=&\frac{1}{2\kappa_4^2}\,\int\,\left[-\frac12e^{12U-\phi}\dd b_2\wedge*\dd b_2-\frac32e^{2U+\frac{\phi}{2}}f_2\wedge*f_2-\frac12e^{6U+\frac32\phi}\dd\tilde{a}_1\wedge*\dd \tilde{a}_1\right.\nonumber\\[2mm]
&&\qquad\qquad\left.-2e^{-6U+\frac{\phi}{2}}\text{D}a_0\wedge*\text{D}a_0-6\,\Qf \, e^{3U-\frac{\phi}{4}}\cF_2\wedge*\cF_2-\Qc\,\tilde{a}_1\wedge\dd b_2\right]\,, \qquad
\end{eqnarray}
where the scalars are to be understood as taking the values in (\ref{AdSsolution}) and we have defined the field strengths 
\begin{equation}
\text{D}a_0\,=\,\dd a_0-3\,a_1\,,\qquad \cF_2\,=\,2\pi\ell_s^2 \, \dd A+b_2\,,\qquad f_2\,=\,\dd a_1+\Qf\,\cF_2\,.
\end{equation}
The kinetic term for $\cF_2$ descends from the (smeared) DBI action for the D6-branes expanded to quadratic order in the fluctuations. The last term is topological but plays an important role in the diagonalization. The first step is to get rid of the would-be mass terms for $b_2$ contained in $f_2$ and $\cF_2$, which is achieved by dualizing $A$ to a new vector $\widetilde{A}$. Let us define $2\pi\ell_s^2 \, \dd A=H_2$ and $\widetilde{H}_2=\dd \widetilde{A}$ and add to the action the total derivative
\begin{equation}
\frac{\Qc}{2\kappa_4^2}\,\int\,H_2\wedge\widetilde{H}_2\,.
\end{equation}
Varying with respect to $H_2$ and substituting back the result we obtain an action that depends on $b_2$ only through derivatives
\begin{equation}
S_{\text{\tiny{vector}}}\,=\,\frac{1}{2\kappa_4^2}\,\int\,\left[-\frac12e^{12U-\phi}\dd b_2\wedge*\dd b_2+\Qc\,\dd b_2\wedge\left(\tilde{a}_1+\widetilde{A}\right)+\dots\right]\,.
\end{equation}
This is now suitable for dualization of the two-form to a scalar. This results in a St\"uckelberg coupling precisely to the combination $\tilde{a}_1+\widetilde{A}$, which in this way becomes massive. The resulting spectrum is displayed in Table~\ref{vectorspectrum}. There are in addition two axionic scalars, $a_0$ and the dual to $b_2$, that are of course massless. It can be seen that the vector corresponding to the dimension $\Delta=4$ operator is switched on in our flows, whereas the one corresponding to the $\Delta=5$ vector operator is turned off by our choice of $b_2=0$.

\begin{table}[h]
\resizebox{0.95\textwidth}{!}{\begin{minipage}{\textwidth}
\begin{center}
\begin{tabular}{ccc}
\hline
\rowcolor{gray}	Mass Eigenstate&$ \qquad m^2L^2$&$\qquad\Delta$\\[6pt]
\hline
$\delta \tilde{a}_1+\delta\widetilde{A}$&$\qquad12$&$\qquad5$\\[6pt]
\rowcolor{gray}	$\delta a_1$&$\qquad6$&$\qquad4$\\[6pt]
$\delta \tilde{a}_1-3\,\delta\widetilde{A}$&$\qquad0$&$\qquad2$\\[6pt]
\hline
\end{tabular}
\end{center}
\caption{\small Spectrum of vectors around the supersymmetric  AdS$_4$  solution.}
\label{vectorspectrum}
\end{minipage} }
\end{table}

\section{Calculation of the Wilson loop}
\label{wilson}
A natural observable to study is the Wilson loop. We perform a simple calculation for a time invariant configuration of two external sources placed at the boundary, separated by a distance $L$, and a string hanging between them in the bulk.
We choose a parametrization on the world-sheet by coordinates $\sigma$ and $\tau$. The string world-sheet in the bulk is then given by an embedding $X^M(\sigma, \tau)$ and the Nambu-Goto action is
\begin{equation}
S_\mt{NG}=\frac{1}{2\pi\ls^2} \int_C\d\sigma \, \d\tau\sqrt{\det_{ab}(G_{MN}\partial_a X^M\partial_bX^N})\,.
\end{equation}
We choose 
\be
t=\tau \sac x=\sigma \sac y=y(\sigma)\equiv y(x)
\ee
and we obtain the following action
\begin{equation}
S_\mt{NG}=\frac{1}{2\pi\ls^2} \int_{-L/2}^{L/2}\int_0^T\d t \, \d x\sqrt{-G_{tt}(G_{xx}+G_{yy}y'^2)}\,.
\end{equation}
Since the problem is  invariant under time translations, the energy
\begin{equation}
H= \frac{\delta L}{\delta(\partial_x y)}\partial_x y-L=\frac{G_{xx}G_{tt}}{\sqrt{-G_{tt}(G_{xx}+G_{yy}y'^2)}}
\end{equation} 
is conserved. Its value can be computed at the turning point of the string, which for convenience is taken to be at $x=0$ and $y=y_0$. Since at that point $y'(0)=0$, we have 
\be
H=-\sqrt{-G_{xx}(0) G_{tt}(0)} \,. 
\ee
This leads  to the following differential equation for $y$
\begin{equation}\label{eq.WLyprime}
y'=\sqrt{\frac{G_{xx}}{G_{yy}}}\sqrt{\frac{G_{xx}G_{tt}-G_{xx}(0) G_{tt}(0)}{G_{xx}(0) G_{tt}(0)}}\,,
\end{equation}
from which we get
\begin{equation}
\label{LLL}
\frac{L}{2}=\int_0^{L/2} \d x=\int_{r_0}^{\infty} \frac{\d y}{y'} \,.
\end{equation}

We saw in Sec.~\ref{scalings} that the dependence of the supergravity equations of motion on the several parameters of the theory can be reduced  to a dependence on a single parameter $\rho$ through the rescalings \eqref{rescresc} and \eqref{redef}. In order to reduce the Nambu-Goto equations of motion to the same dependence we must further rescale the $x$-coordinate as in Eqn.~\eqq{x_scaling}. Under these circumstances the Nambu-Goto action scales homogeneously as 
$S_\mt{NG} = \Qf \, \overline S_\mt{NG}$. One immediate consequence of \eqref{x_scaling} is that the distance between the endpoints of the strings satisfies the same scaling property
\be\label{eq.xscaling}
L(\lambda,\nc,\nf,\nq) = \Qf \left( \frac{\Qc}{\Qf^5}\right)^{1/2} \overline L(\rho) \ ,
\ee
and the energy of the system, which is evaluated by calculating the on-shell action divided by the integral over time, can be expressed in terms of dimensionless variables as
\begin{equation}
E(\lambda,\nc,\nf,\nq)= \frac{S}{T} =\Qf \frac{\overline S}{T} = \frac{2\, \Qf}{2\pi\ls^2} \int_{\overline y_0}^\infty \sqrt{-\overline{G}_{tt}(\overline{G}_{xx}+\overline{G}_{yy}\overline{y}'^2)}\,\frac{\d \overline y}{\overline{y}'} \equiv \frac{\Qf}{2\pi\ls^2} \, \overline E(\rho)\,.
\end{equation}
The above quantity is divergent. The interpretation of this divergence is that we are really computing the energy of a pair of quarks including their large self-energy. Since we are only interested in the potential energy of the two sources, we subtract the energy of each individual quark, corresponding to a string stretching all the way from $\overline y=0$ to $\overline y=\infty$. The final result is then given by
\begin{equation}
\overline E=2\left(\int_{\overline y_0}^\infty \sqrt{-\overline{G}_{tt}(\overline{G}_{xx}+\overline{G}_{yy}\overline{y}'^2)}\,\frac{\d \overline y}{\overline{y}'} -\int_0^\infty \sqrt{-\overline{G}_{tt}\overline{G}_{yy}}\, \d \overline y \right)\,.
\label{E10}
\end{equation}

Applying the above procedure to the D2-brane solution we obtain
\begin{equation}
\overline L\sim \overline y_0^{-3/2}\,, \quad \overline E\sim \overline y_0\,.
\end{equation}
In the case of  AdS$_4$, in the coordinate system used here, we get
\begin{equation}
\overline L\sim \overline y_0^{-5/4}\,, \quad \overline E\sim \overline y_0^{5/4}\,.
\end{equation}
More precisely, using \eqq{E10} and the rescaled version of \eqq{LLL} we have that the product $\overline E \cdot \overline L$ in the AdS$_4$ region is a constant given by \eqq{ELads}.
Finally, for the HVL geometry  the results are
\begin{equation}
\overline L\sim \overline y_0^{-5/6}\,, \quad \overline E\sim \overline y_0^5\,.
\end{equation}
In Fig.~\ref{fig:WL}, we have plotted the potential energy $\overline E$ in units of $1/\overline L$  against the distance between the two sources, $\overline L$. We see that for small $\overline L$, the Wilson loop is probing only the UV part of the geometry and thus the energy scales as in the  D2-brane solution. For large $\overline L$, the string penetrates deeper in the bulk, probing the IR region, and thus the scaling is given by the HVL solution. At intermediate values of $\overline L$, we see an AdS region building up.

\end{document}